\documentclass[fleqn,usenatbib,useAMS]{mnras}
\usepackage[utf8]{inputenc}

\usepackage[T1]{fontenc}
\usepackage{graphicx}
\usepackage{float}
\DeclareGraphicsExtensions{.pdf,.png}
\usepackage[dvipsnames]{xcolor}
\usepackage{amsmath}
\usepackage{amssymb}
\usepackage{nicefrac}
\usepackage{comment}
\usepackage[normalem]{ulem}
\usepackage{footnote}
\usepackage{dsfont}
\usepackage{enumitem}
\usepackage{bbold}
\usepackage{bm}

\DeclareFontFamily{OT1}{pzc}{}
\DeclareFontShape{OT1}{pzc}{m}{it}{<-> s * [1.10] pzcmi7t}{}
\DeclareMathAlphabet{\mathpzc}{OT1}{pzc}{m}{it}

\definecolor{blue}{rgb}{0,0,1}
\definecolor{darkergreen}{rgb}{0,0.5,0}

\defcitealias{Planck2018PNG}{Planck Collaboration (2019)}

\newcommand{\eg}{e.g$.$ }
\newcommand{\Eg}{E.g$.$ }
\newcommand{\egnospace}{e.g$.$}

\newcommand{\Ie}{I.e.\ }
\newcommand{\ie}{i.e$.$ }
\newcommand{\cf}{cf$.$ }
\newcommand{\cfnospace}{cf$.$}

\newcommand{\wrt}{wrt$.$ }
\newcommand{\dd}{\ensuremath{\mathrm{d}}}

\newcommand{\figref}[1]{Figure \ref{fi:#1}}

\newcommand{\eqnref}[1]{Equation \ref{eq:#1}}
\newcommand{\eqnrefnospace}[1]{Equation \ref{eq:#1}}
\newcommand{\Eqnref}[1]{Equation \ref{eq:#1}}

\newcommand{\appref}[1]{Appendix \ref{app:#1}}
\newcommand{\apprefnospace}[1]{Appendix \ref{app:#1}}

\newcommand{\secref}[1]{Section \ref{sec:#1}}
\newcommand{\secrefnospace}[1]{Section \ref{sec:#1}}
\newcommand{\Secref}[1]{Section \ref{sec:#1}}

\allowdisplaybreaks

\title{Bye binormal: analysing the joint PDF of galaxy density and weak lensing convergence}

\author[]{Oliver Friedrich$^{1,2}$, Lina Castiblanco$^{3,4}$, Anik Halder$^{1,5,6}$, Cora Uhlemann$^{3,4}$\\
$^{1}$ Universitäts-Sternwarte, Fakultät für Physik, Ludwig-Maximilians Universität München, Scheinerstr. 1, 81679 München, Germany\\
$^2$ Excellence Cluster ORIGINS, Boltzmannstr.\ 2, 85748 Garching, Germany\\
$^3$ Fakultät für Physik, Universität Bielefeld, Postfach 100131, 33501 Bielefeld, Germany\\
$^{4}$ School of Mathematics, Statistics and Physics, Newcastle University, Herschel Building, NE1 7RU Newcastle-upon-Tyne, U.K.\\
$^5$ Institute of Astronomy and Kavli Institute for Cosmology, University of Cambridge, Madingley Road, Cambridge CB3 0HA, UK\\
$^6$ Jesus College, Jesus Lane, Cambridge, CB5 8BL, UK
}

\begin{document}

\maketitle

\begin{abstract}
At any given scale, 3$\times$2-point statistics extract only three numbers from the joint distribution of the cosmic matter density and galaxy density fluctuations: their variances and their covariance. It is well known that the full shape of the PDF of those fluctuations contains significantly more information than can be accessed through these three numbers. But the study of the PDF of cosmic density fluctuations in real observational data is still in its infancy. Here we present \verb|CosMomentum|, a public software toolkit for calculating theoretical predictions for the full shape of the joint distribution of a line-of-sight projected tracer density and the gravitational lensing convergence. We demonstrate that an analysis of this full shape of the PDF can indeed disentangle complicated tracer bias and stochasticity relations from signatures of cosmic structure growth. Our paper also provides back-drop for an upcoming follow-up study, which prepares PDF analyses for application to observational data by  incorporating the impact of realistic weak lensing systematics. 
\end{abstract}

\section{Introduction}

One of the major questions asked by today's cosmological studies of the large-scale structure of the Universe is in fact a very simple one: Do we understand gravitational collapse on cosmological distance and time scales? The archetypal experiment to answer this question is to measure the amplitude of density fluctuations in the early Universe and test whether -- starting from those fluctuations -- the cosmological standard model can predict the amplitude of fluctuations today. This has \eg been done by comparing cosmological parameter constraints obtained from the power spectrum of temperature and polarisation of the cosmic microwave background \citep[CMB, \egnospace][]{Planck2018cosmo}  to the power spectrum of the galaxy density field and the cosmic gravitational lensing field in the late Universe \citep[\egnospace][]{Hildebrandt2017, DES2018, DES:2021wwk, 2023PhRvD.108l3517M, 2025arXiv250319441W}. While the most recent instalments of this comparison find no significant deviation between large-scale structure data and the cosmological standard model, recent spectroscopic analyses \citep{2025JCAP...02..021A, 2025arXiv250314738D} indicate that such a deviation does indeed exist. But both statistical and systematic uncertainties limit our ability to pin down such beyond-standard-model physics with photometric surveys of structure growth.

This limitation is partly due to the fact that we primarily test our understanding of gravitational collapse through the analysis of density fluctuation variance as a function of time and scale (2-point statistics). \Ie at any given distance scale and time, we only extract 3 numbers from the joint probability distribution functions of the cosmic matter density and galaxy density field: their two variances and their covariance. And these numbers vary so smoothly with both time and scale that such analyses lead to featureless signals (especially in photometric surveys) that can be too poor to distinguish between cosmological signatures and signatures of poorly understood survey systematics and astrophysics (baryonic physics). This is exacerbated by the fact that non-linear evolution of the cosmic density field actually removes information from the 2nd moments of density fluctuations and stores that information in higher order correlation functions, where it cannot be accessed by 2-point statistics.

To overcome these limitations, future analyses of the large-scale structure of the Universe must uncover a more complete view of the cosmic density field. This may \eg be achieved by studying higher-order moments of the density field \citep{Petri:2015ura,Peel:2018aly,DES:2021lsy,DES:2023qwe}, higher-order correlation functions \citep{Halder:2021itp, Halder2022, Halder_2023, Gong2023, Burger:2023qef}, or complementary statistics such as marked correlation functions \citep{Aviles:2019fli,Armijo:2023sld,Lai:2023dzp}, Minkowski functionals \citep{Grewal:2022qyf,Armijo:2024ujo,Liu:2025haj}, or peak statistics \citep{Martinet:2017rqp,Harnois-Deraps:2020pvj,DES:2021epj,Marques:2023bnr}. Another promising route to analyse the large-scale structure beyond its 2-point statistics is the study of the full shape of the probability distribution function (PDF) of density fluctuations. Promising forecasts of higher-order weak lensing statistics for Euclid \citep{EuclidHOWLSKP1} show that this PDF contains significant cosmological information, and combining it with standard 2-point statistics can significantly tighten cosmological constraints \citep[as demonstrated in Figure~43 in][]{EuclidOverview}.
It has long been known that this shape can be well understood analytically (see \eg \citealt{Bernardeau1994, Bernardeau1995, Bernardeau2000, Valageas2002}). More recent theoretical advances have been made for the PDF of matter density \citep{Bernardeau2014, Bernardeau2015, Uhlemann2016, Ivanov2019} and its sensitivity to cosmology and fundamental physics \citep{Codis2016a,Uhlemann2018b,Uhlemann2020, Friedrich2020,Cataneo2022,Coulton2024}. Those results can be linked to the PDFs of weak gravitational lensing and galaxy clustering fields such as the convergence \citep{Barthelemy2020nulling,Barthelemy2020postBorn,Boyle2021,Barthelemy2021jointnulling,Barthelemy2024,Castiblanco2024,Sreekanth2024} and aperture mass \citep{Barthelemy2021apm}, and counts-in-cells of biased tracers \citep{Uhlemann2018a,Leicht2019,Friedrich2021b,Britt2024,Gould2025}. Another advantage of one-point PDFs is that their analytical covariances are more tractable as they can be determined from the joint two-point PDFs \citep{Uhlemann2023PDFcov}.
Nearest Neighbour Statistics \citep{Banerjee2021kNN} are closely related to density PDFs and similarly can be extended to capture cross-correlations between a tracer and a field \citep{Banerjee2023kNNcross}. Those statistics organise information in a slightly different way, while density PDFs scan densities at a given set of smoothing scales, $k$NN statistics scan scales at a given number of nearest neighbours.

In observational data, analyses have so far focused on studying the PDF of galaxy density fluctuations \citep[\egnospace][]{Efstathiou1995, Yang2011, Salvador2019, Repp2020}, while the PDF of gravitational lensing convergence (which probes the total matter density field) has only been analysed based on a simplified shifted log-normal model \citep{Hilbert2011} by \citet{Clerkin2017}. A breakthrough in comparing our theoretical understanding of the full matter density PDF with observations was made through the program of density-split statistics \citep{Gruen2016, Friedrich2018, Gruen2018, Brouwer2018, Burger2020, Burger2021, Paillas2021}. By studying galaxy counts-in- and gravitational lensing-around-cells \citep[or redshift space distortions in split densities, as considered by][]{Paillas2021} these studies were able to extract measurements from galaxy surveys that indirectly trace the shape of the density PDF. And based on these measurements, as well as the aforementioned theory \citet{Friedrich2018} and \citet{Gruen2018} derived for the first time cosmological information from the PDF of density fluctuations that approaches the precision of state-of-the-art 2-point analyses. Ultimately, the main role of PDF analyses may however not be to compete with 2-point statistics, but to serve as an anchor to pin down small scale physics: the PDF is extremely versatile in distinguishing aspects of the large-scale structure that leave very degenerate signatures in 2-point correlation functions \citep{Friedrich2018, Friedrich2021b}.

In this paper we take the PDF program one step further and develop the techniques to analyse the full shape of the joint PDF of galaxy density fluctuations and weak gravitational lensing convergence. In particular, we prepare for an analysis of fluctuations in the density of MagLim galaxies from the Dark Energy Survey \citep[DES,][]{DES:2021wwk,DES:2021bpo} 
and the convergence of the weak gravitational lensing of galaxies probed by a Stage-IV survey like the Euclid space telescope \citep{EuclidOverview,EuclidHOWLSKP1} and Rubin Observatory Legacy Survey of Space and Time \citep[LSST,][]{LSST_Ivezic_2019}. Our presentation is set out as follows: \Secref{model} presents the details of our theoretical modelling. In \secref{validation} we describe different sets of simulated data and use these data to validate our theoretical predictions. \Secref{discussion} closes with a number of conclusions, an outlook on the potential of our method for current and future galaxy surveys, and a discussion of open issues. We accompany our analysis with the publicly available numerical toolkit \verb|CosMomentum|\footnote{\url{https://github.com/OliverFHD/CosMomentum}}, which we hope will foster further development and a wider application of the methods developed here.

\section{Theory}
\label{sec:model}

\begin{figure}
    \includegraphics[width=0.5\textwidth]{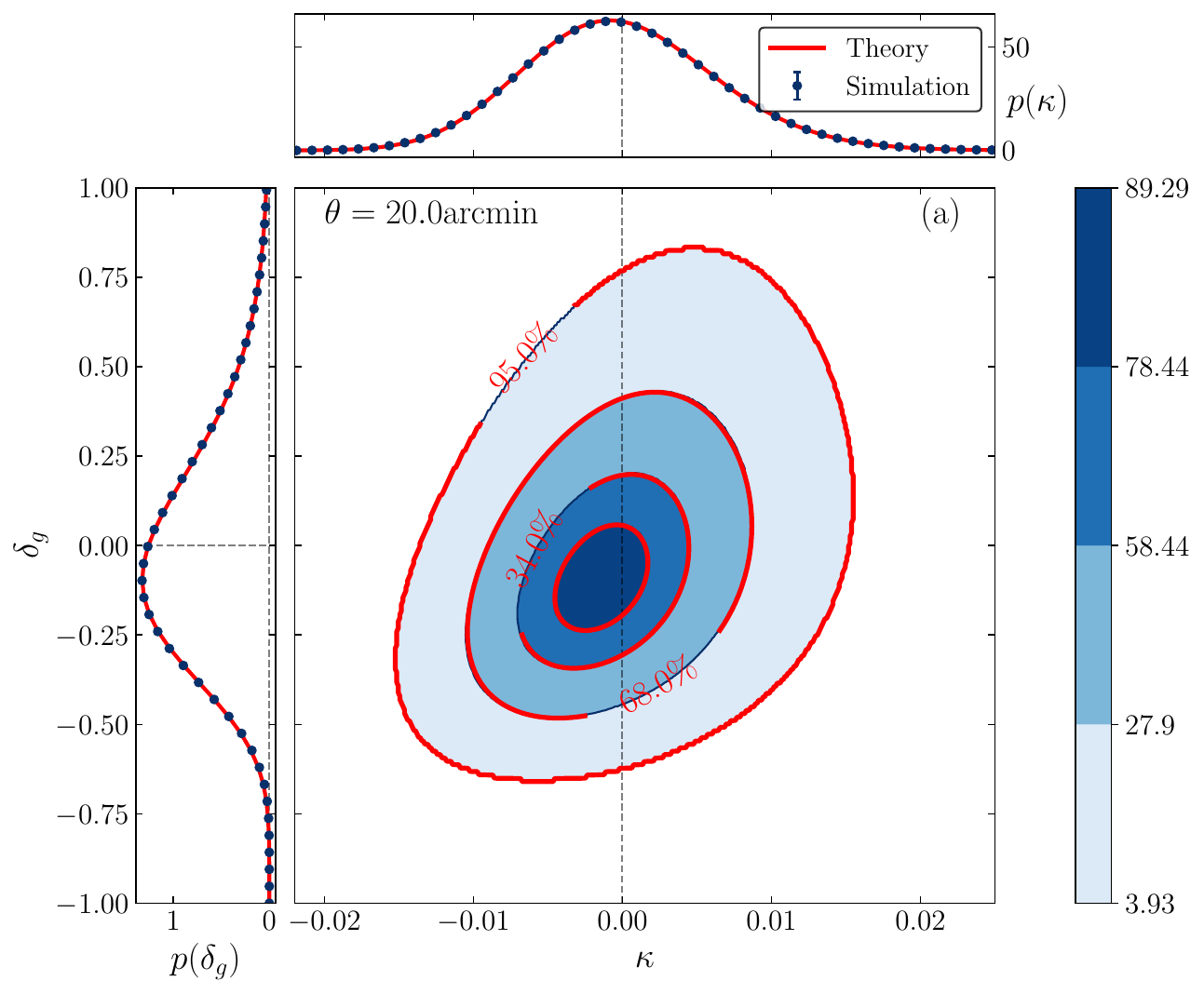}
    \includegraphics[width=0.5\textwidth]{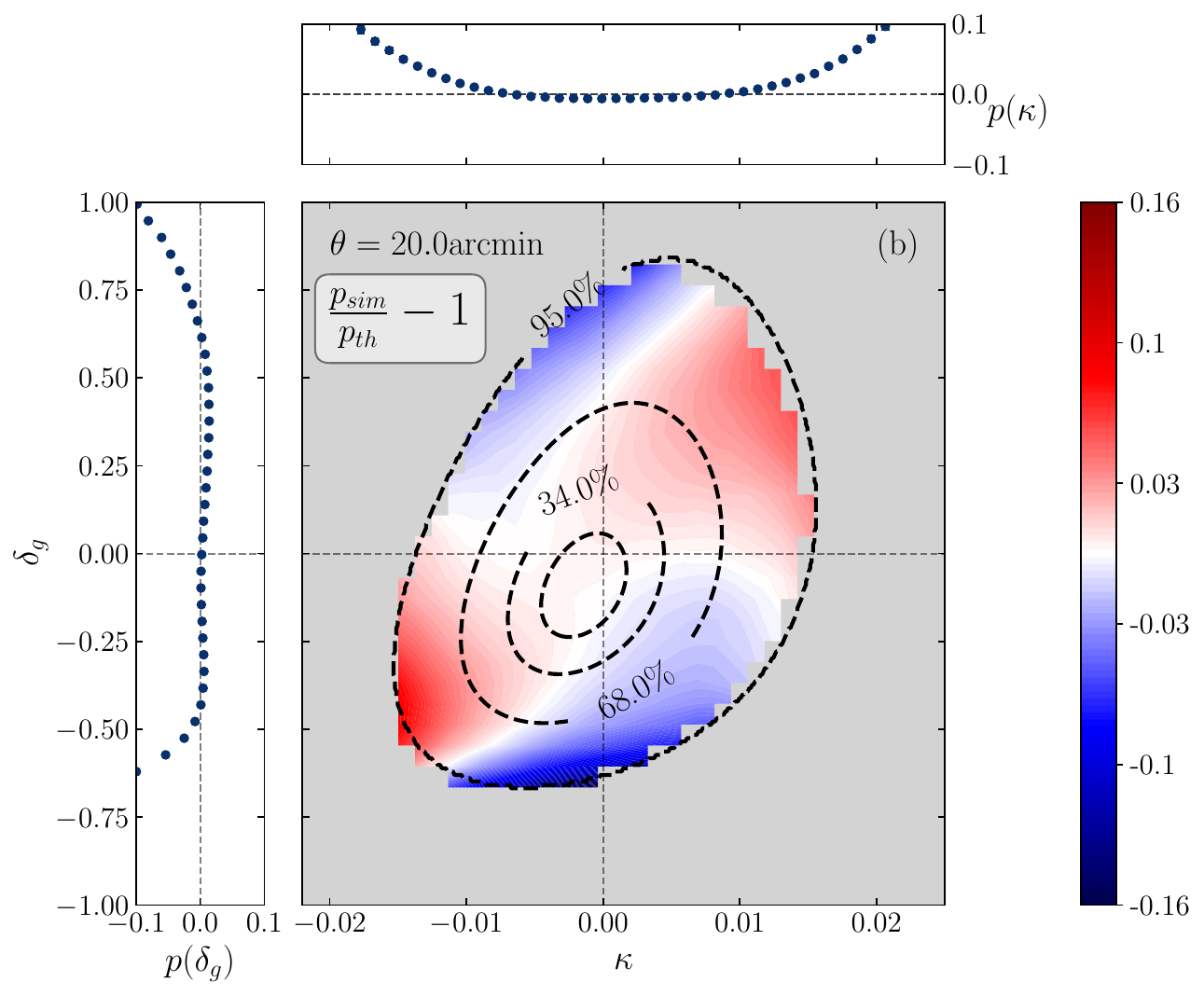}
    \caption{\textbf{(a).} Central panel: Fiducial theoretical prediction for the joint PDF $p(\delta_g, \kappa)$  shown as red contours, enclosing  $12\%$, $34\%$,$ 68\%$ and $95\%$ of the total probability, where $\delta_g$ is density contrast of DES-like MagLim galaxies and $\kappa$ is the Stage-IV-like weak lensing convergence- both averaged over $20$arcmin filters on the sky. For the weak lensing convergence $\kappa$ we include the effect of shape noise from intrinsic galaxy shapes. For comparison, blue contours show the joint PDF measured from mock catalogues based on the T17 simulations. Upper panel: Predicted (red solid line) and measure (blue points) convergence PDF $p(\kappa)$. Left panel: Predicted (red solid line) and measured (blue points) galaxy number count PDF $p(N_g)$. $p(\kappa)$ and $p(N_g)$ are obtained by marginalizing the joint PDF over $N_g$ and $\kappa$, respectively. \textbf{(b).} Residuals between the measured and predicted joint and marginal PDFs. We recommend excluding the grey outer regions (beyond the $95\%$ probability region) of the $\delta_g-\kappa$ plane in forecasts and data analyses due to limited model accuracy and the breakdown of the Gaussian likelihood approximation in those areas. Dashed black lines represent the theory predictions for $p(\delta_g, \kappa)$ to highlight areas of high probability.}
  \label{fi:fiducial_PDF}
\end{figure}

In this paper we are interested in the joint PDF of projected galaxy density contrast $\delta_g$ and weak lensing convergence $\kappa$ on the sky. Let us initially ignore the issue of galaxy bias and stochasticity and consider the projected matter density contrast $\delta_m$ instead of $\delta_g$. Both $\delta_m$ and $\kappa$ can be considered as line-of-sight projections of the 3-dimensional density contrast $\delta_{3\mathrm{D}}$. Identifying points on the sky by unit vectors $\mathbf{\hat n}$ on the sphere and assuming a spatially flat Universe, this can be written as  \citep{BartelmannSchneider2001}

\begin{align}
    \label{eq:def-lens-deltam}
    \delta_m(\mathbf{\hat n}) =&\ \int \dd \chi\ w_m(\chi)\ \delta_{3\mathrm{D}}(\chi\cdot \mathbf{\hat n},\eta_0 - \chi/c)\,,\\
    \label{eq:def-convergence}
    \kappa(\mathbf{\hat n}) =&\ \int \dd \chi\ w_l(\chi)\ \delta_{3\mathrm{D}}(\chi\cdot \mathbf{\hat n}, \eta_0 - \chi/c)\ .
\end{align}
Here $\chi$ is co-moving distance, $\eta_0$ is today's conformal time, and $w_m$ and $w_l$ are the line-of-sight projection kernels corresponding to the two fields. The kernel $w_m(\chi)$ is given in terms of the redshift distribution $n_g(z)$ of the tracer (lens) galaxies as
\begin{equation}
    w_m(\chi) = n_g(z(\chi))\  \frac{\dd z}{\dd \chi}(\chi)\,,
\end{equation}
and the weak lensing kernel is given by the source galaxy distribution $n_s(z_s)$
\begin{align}
\label{eq:lensing_kernel}
w_l(\chi) &= \int dz_s\,n_s(z_s) w_{l,z_s}(z(\chi))\,,\\
\label{eq:lensing_kernel_zs}
w_{l,z_s}(z) &= \frac{3H_{0}^{2}\Omega_{m}}{2c^2} \frac{\chi(z)\big[\chi(z_s)-\chi(z)\big]}{\chi(z_s)}(1+z) \Theta(z_s-z)\,,
\end{align}

where $\Omega_m$ is today's total matter density in units of the critical density, $H_0$ is today's Hubble expansion rate, $c$ is the speed of light and $\Theta$ is the Heaviside function. 

\begin{figure}
    \centering
    \includegraphics[width=0.95\linewidth]{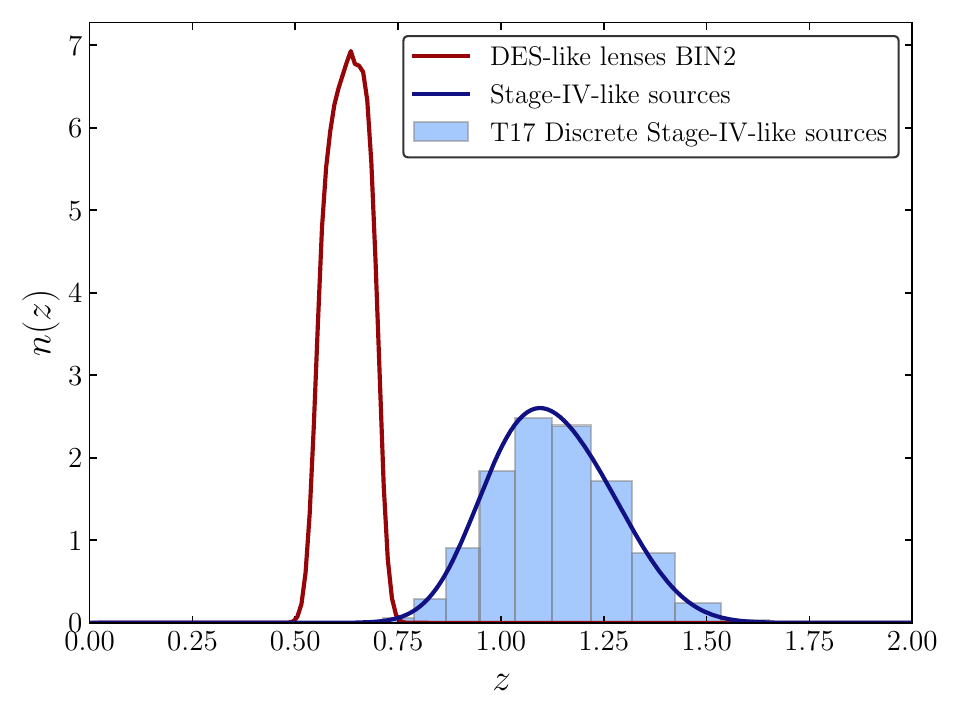}
    \caption{Redshift distributions of the simulated DES-like lenses (dark red solid line) and Stage-IV-like sources (dark blue solid line). The histogram represents the discrete Stage-IV-like source redshift distribution used to create the source catalogue from the \citet{Takahashi2017} (T17) simulation suite.  }
    \label{fig:redshift_distributions}
\end{figure}

For our work, we chose a Stage-IV like background source sample peaking at redshift $z_s\approx 1.1$ and a
DES-like lens sample in the foreground with a peak at $z_l\approx 0.6$ (corresponding to redshift bin 2). This ensures a good overlap between the lens redshift distribution and the weak lensing kernel while minimising the overlap in the redshift distributions to mitigate strong effects from intrinsic alignments. We illustrate the redshift distributions of our lenses and sources in Figure~\ref{fig:redshift_distributions}.  

In the following notation, let us fix some angular scale $\theta$ and consider averages of $\delta_m$ and $\kappa$ over circles of radius $\theta$, \ie
\begin{align}
    \delta_{m,\theta}(\mathbf{\hat n}) =&\ \frac{1}{2\pi(1-\cos\theta)} \underset{\mathbf{\hat n}\cdot\mathbf{\hat n}'<\cos\theta}{\int} \dd\Omega'\ \delta_{m}(\mathbf{\hat n}')\,, \\
    \kappa_{\theta}(\mathbf{\hat n}) =&\ \frac{1}{2\pi(1-\cos\theta)} \underset{\mathbf{\hat n}\cdot\mathbf{\hat n}'<\cos\theta}{\int} \dd\Omega'\ \kappa(\mathbf{\hat n}')\ .
\end{align}
We want to model the probability distribution function (PDF) of these averaged quantities. Assuming that the Universe is statistically isotropic, this PDF will not depend on the location $\mathbf{\hat n}$ and we can simply denote it by $p(\delta_{m,\theta}; \kappa_{\theta})$.

To model the PDF $p(\delta_{m,\theta}; \kappa_{\theta})$ theoretically it is useful to consider also the cumulant generating function (CGF) of the two fields. First, their moment generating function is given by the power series
\begin{equation}
    \psi_\theta(\lambda_\delta, \lambda_\kappa) = \sum_{i,j=0}^\infty \frac{\langle \delta_{m,\theta}^i\ \kappa_{\theta}^j \rangle}{i!\ j!}\ \lambda_\delta^i\ \lambda_\kappa^j\,,
\end{equation}
and the CGF is defined as its logarithm, \ie
\begin{align}
    \varphi_\theta(\lambda_\delta, \lambda_\kappa) =&\ \ln\psi_\theta(\lambda_\delta, \lambda_\kappa) \nonumber \\
    \equiv &\ \sum_{i,j=0}^\infty \frac{\langle \delta_{m,\theta}^i\ \kappa_{\theta}^j \rangle_c}{i!\ j!}\ \lambda_\delta^i\ \lambda_\kappa^j\ .
\end{align}
Here, the last line serves as a definition of the so called cumulants (or connected moments) $\langle \dots \rangle_c$. Those are the parts of the moments that would vanish if $\delta_{m,\theta}$ and $\kappa_{\theta}$ had a multivariate Gaussian distribution \citep{Bernardeau2015}.

The pair PDF-CGF is somewhat analogous to the pair of real space and Fourier space correlation functions in 2-point statistics. In data, it is more straightforward to measure the PDF and the real space correlation function. But in cosmological theory it is more straightforward to predict the CGF \citep[which could also be an observable][]{Boyle2023CGF} and the Fourier space correlation function (the power spectrum). While power spectrum and real space correlation function are connected via a Fourier transform, the PDF and CGF are related through a Laplace transform
\begin{align}
\label{eq:CGF_as_expectation}
    &\ e^{\varphi_\theta(\lambda_\delta, \lambda_\kappa)}\nonumber \\
    =&\ \langle e^{\lambda_\delta \delta_{m,\theta} +  \lambda_\kappa \kappa_{\theta}} \rangle \nonumber \\
    =& \int \dd \delta_{m,\theta}\  \dd \kappa_{\theta}\ p(\delta_{m,\theta}; \kappa_{\theta}) \ e^{\lambda_\delta \delta_{m,\theta} +  \lambda_\kappa \kappa_{\theta}}\ \,.
\end{align}
Furthermore, following \citet{Bernardeau2000} \citep[see also][]{Friedrich2018, Barthelemy2020nulling, Boyle2021} the joint projected CGF $\varphi_\theta(\lambda_\delta, \lambda_\kappa)$ can be approximated in terms of the CGF of 3-dimensional matter density contrast in cylinders of radius $R$ and length $L$, $\varphi_{\mathrm{cy}, R, L}(\lambda, \eta)$, as
\begin{align}
\label{eq:CGF_Limber}
    &\ \varphi_\theta(\lambda_\delta, \lambda_\kappa)\nonumber \\ \approx &\ \int \dd \chi\ \lim_{L\rightarrow \infty} \frac{\varphi_{\mathrm{cy}, \chi\theta, L}\left(L\lbrace w_m(\chi)\lambda_\delta + w_l(\chi) \lambda_\kappa \rbrace, \eta_0 - \chi/c\right)}{L}\ .
\end{align}
This can be considered as the equivalent of the Limber approximation \citep{Limber1953} of the line-of-sight projected power spectra for the line-of-sight projected CGF.

\begin{figure} 
    \includegraphics[width=0.45\textwidth]{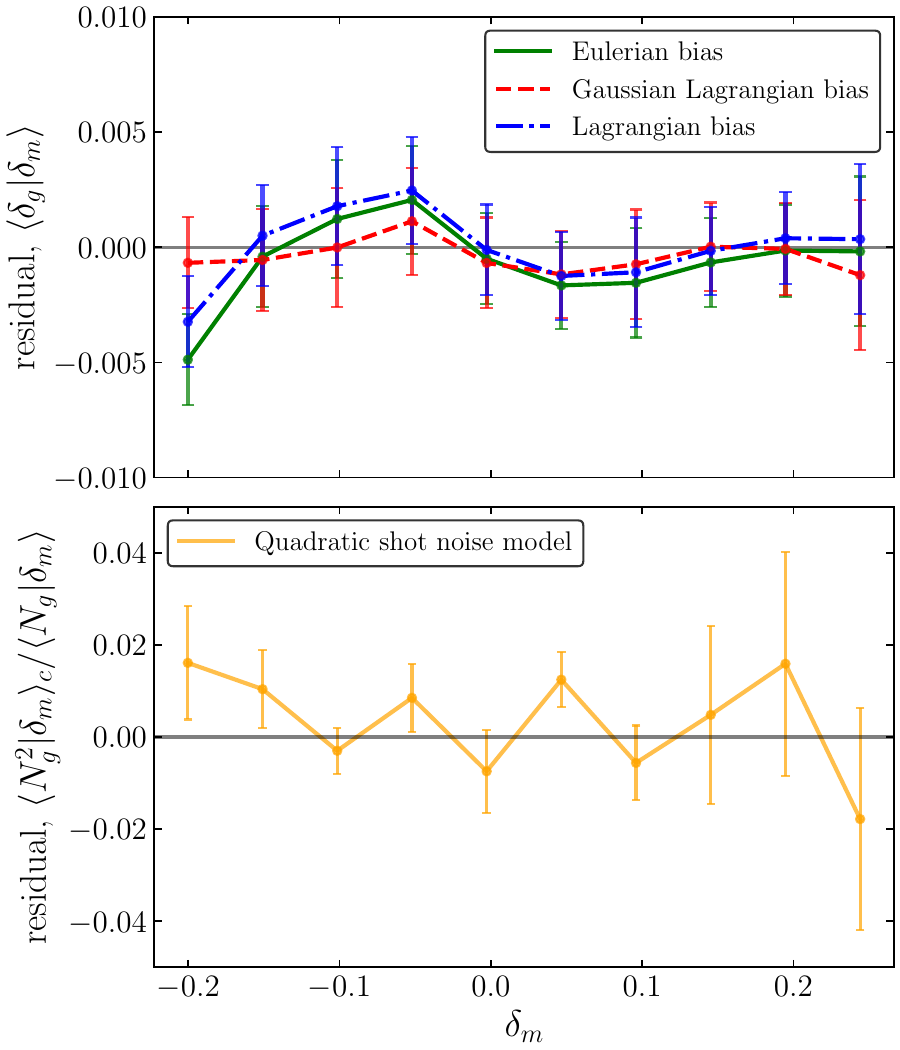}\caption{Upper pannel: Comparison of the residuals between the measured and fitted conditional mean of the number count, or equivalently density $\langle \delta_g|\delta_m\rangle$, using the Eulerian (solid green line), Gaussian Lagrangian bias expansions (dashed red line) and  Lagrangian (dash-dotted blue line). Lower panel: Residual between the measured and fitted conditional variance, using a quadratic shot noise model. Error bars represent the error of the mean over 6 non-overlapping circular footprints of one \citet{Takahashi2017} (T17) realisation.}
  \label{fig:Bias_residual}
\end{figure}

\begin{figure*}
    \centering
    \includegraphics[width=0.5\textwidth]{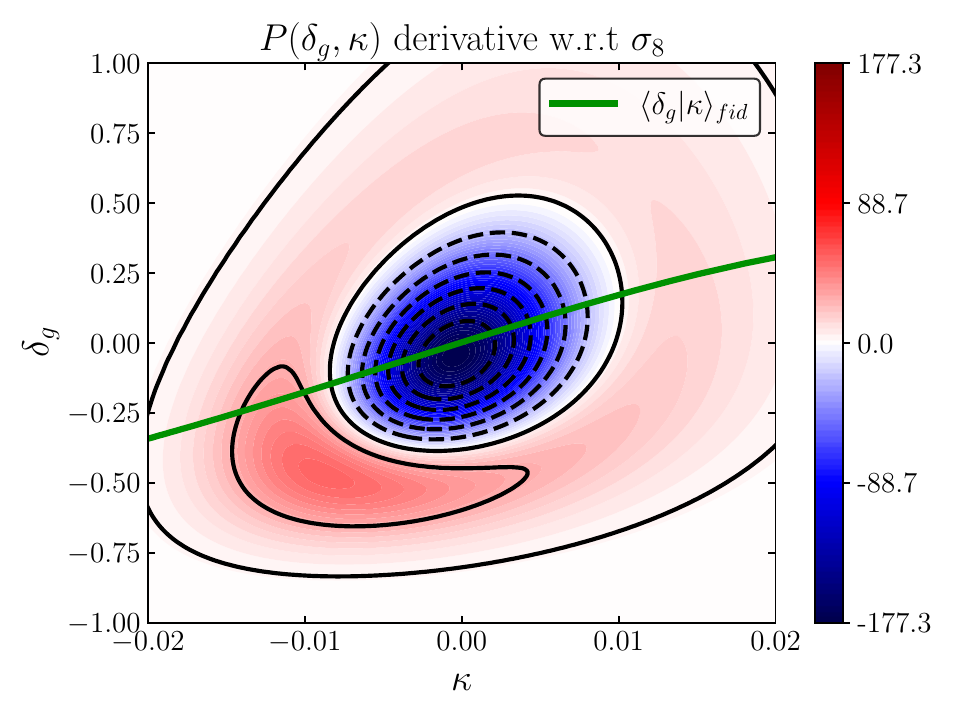}\includegraphics[width=0.5\textwidth]{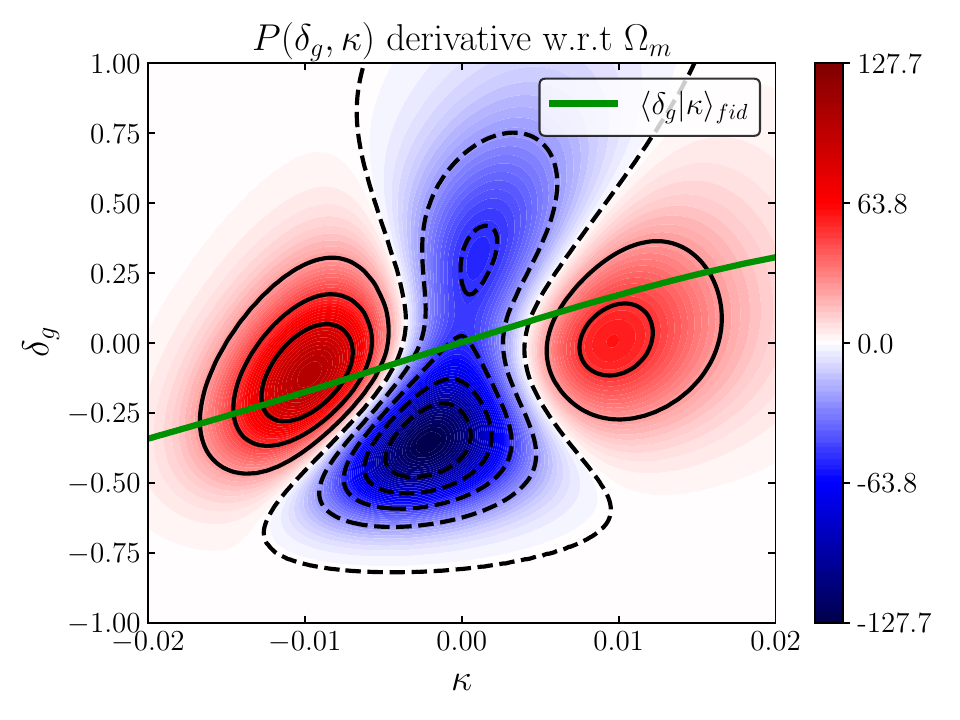}
    
    \includegraphics[width=0.5\textwidth]{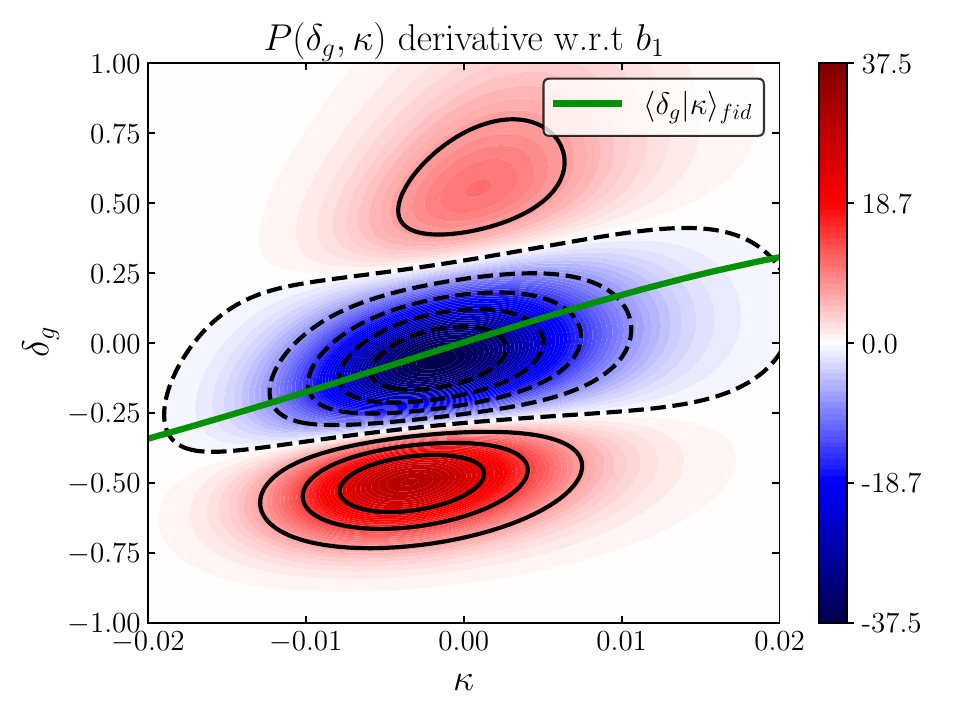}\includegraphics[width=0.5\textwidth]{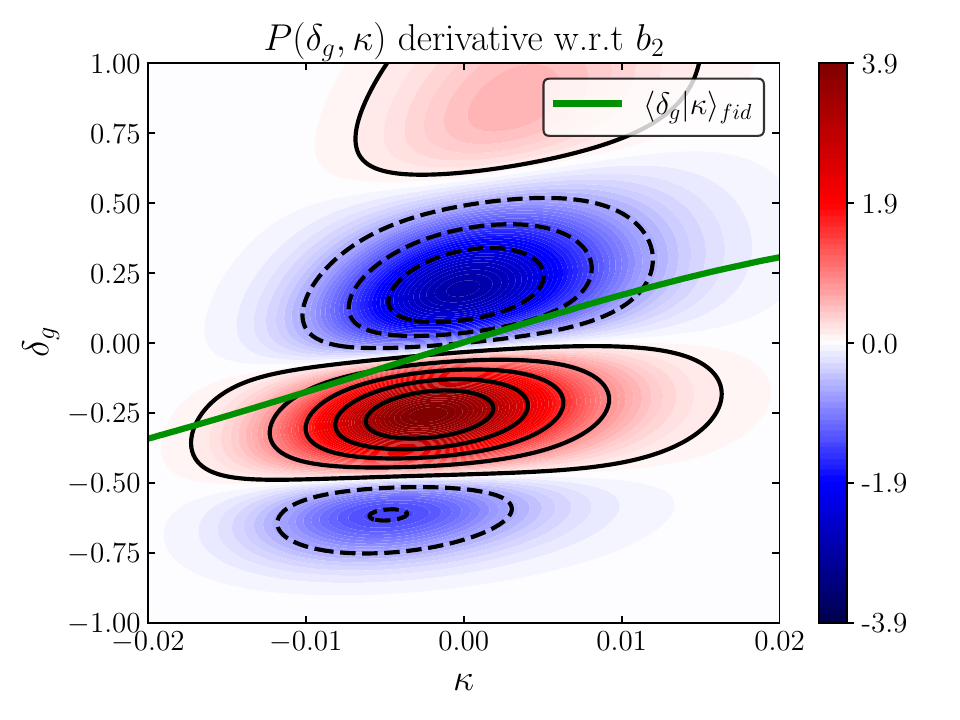}
    
    \includegraphics[width=0.5\textwidth]{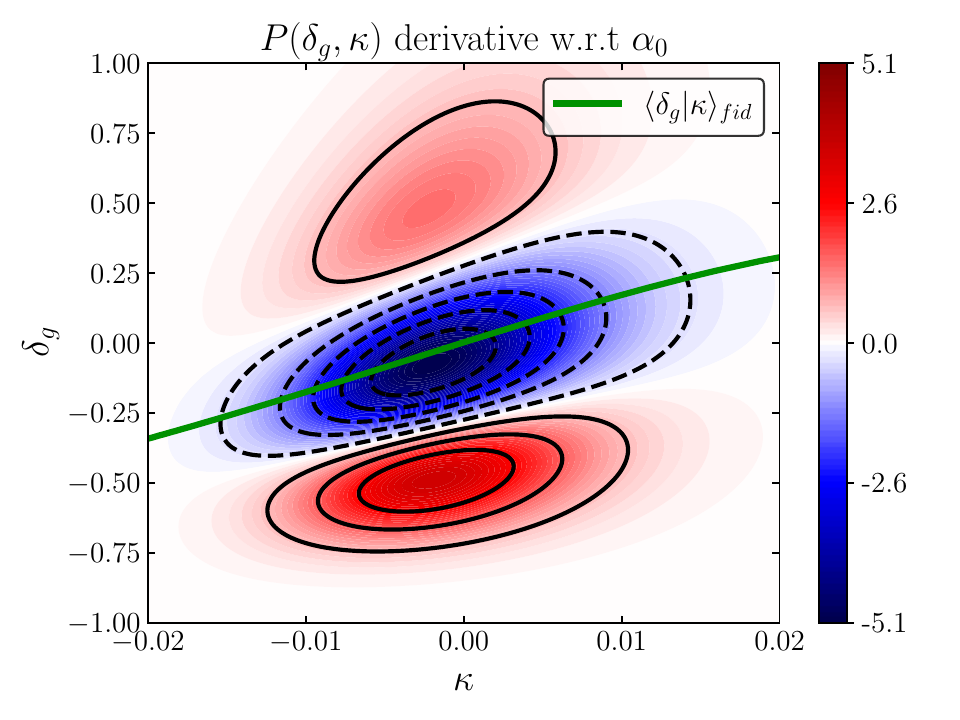}\includegraphics[width=0.5\textwidth]{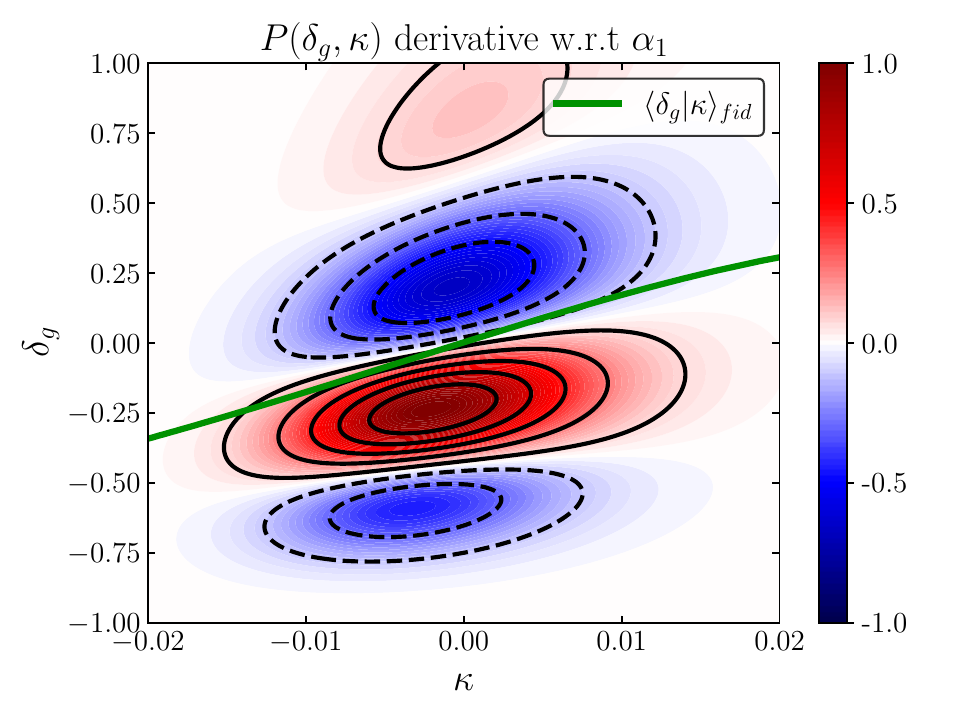}
\caption{Colour maps and black contours (positive/negative value as solid/dashed lines) show the derivatives of $p(\delta_g, \kappa)$ \wrt cosmological parameters $\sigma_8$ and $\Omega_m$ (upper panel), galaxy bias parameters $b_1^E$ and $b_2^E$ (middle panel) and stochasticity parameters $\alpha_0$ and $\alpha_1$ (lower panel). The green line indicates the conditional mean galaxy density contrast given the weak lensing convergence.}
  \label{fi:derivatives}
\end{figure*}

\Eqnref{CGF_as_expectation} can then be inverted to compute the PDF $p(\delta_{m,\theta}; \kappa_{\theta})$ from the CGF $\varphi_\theta(\lambda_\delta, \lambda_\kappa)$ with an inverse Laplace transformation. Choosing the integration paths of that transformation along the imaginary axes of $\lambda_\delta$ and $\lambda_\kappa$ leads to
\begin{align}
\label{eq:inv_laplace}
    &\ p(\delta_{m,\theta}; \kappa_{\theta}) = \nonumber \\
    & \int \frac{\dd \lambda_\delta\  \dd \lambda_\kappa}{(2\pi)^2} \ \exp(-i\lambda_\delta \delta_{m,\theta}  -i\lambda_\kappa \kappa_{\theta}) \exp(\varphi_\theta(i\lambda_\delta, i\lambda_\kappa))\ .
\end{align}
Note that this is simply a 2D Fourier transform of the function $\exp(\varphi_\theta(i\lambda_\delta, i\lambda_\kappa))$. Hence, fast-Fourier-transform (FFT) is a promising technique to evaluate \eqnref{inv_laplace} quickly. 

So far (\cf \eqnrefnospace{CGF_Limber}) we have only considered the joint PDF of projected matter density contrast and lensing convergence, $p(\delta_m, \kappa)$. We now want to replace the unobservable $\delta_m$ with the number count $N_g$ of a galaxy sample with the same line-of-sight projection kernel. To do so, let us first consider the joint PDF of all three variables, which we decompose as
\begin{equation}
\label{eq:3PDF}
    p(N_g, \delta_m, \kappa) = p(N_g|\delta_m)\ p(\delta_m, \kappa)\ .
\end{equation}
Here $p(N_g|\delta_m)$ is the conditional PDF of finding a galaxy number count $N_g$ in our aperture when the matter density contrast in the same aperture is $\delta_m$. This conditional PDF includes the conditional expectation value capturing galaxy bias
\begin{equation}
\label{eq:bias}
    \langle N_g | \delta_m \rangle = \bar{N}_g (1+ \langle \delta_g | \delta_m \rangle)\,,
\end{equation}
where $\bar N_g$ is the mean galaxy count averaged over all apertures on the sky, and the stochasticity (or shot noise) of the galaxy density field around this deterministic relationship. The factorisation of the PDF in \eqnref{3PDF} assumes that this stochasticity is completely determined by $\delta_m$ in the aperture and is otherwise independent of $\kappa$. The PDF of $N_g$ and $\kappa$ is then obtained by integrating out $\delta_m$,
\begin{equation}
\label{eq:jointPDF}
p(N_g, \kappa) = \int \dd\delta_m\ p(N_g|\delta_m)\ p(\delta_m, \kappa)\ .    
\end{equation}

We describe models for the deterministic and stochastic parts of $p(N_g|\delta_m)$ in \secref{cylinder_CGF_with_bias}~and~\secref{shot-noise}, respectively. \Secref{shape_noise} illustrates how to incorporate galaxy shape noise which affects the weak lensing convergence in the joint PDF. 

The remainder of this section will detail analytical and numerical recipes to evaluate key parts of the above equations. \Secref{cylinder_CGF} describes how to model the CGF of matter density contrast in cylinders. \Secref{practical_CGFevaluation}~and~\Secref{invLaplace} describes how to efficiently perform the line-of-sight projection of the CGF and the inversion of the Laplace transform that relates the PDF and CGF of the projected fields.

\subsection{Galaxy bias}
\label{sec:cylinder_CGF_with_bias}
The galaxy density contrast $\delta_g$ inside a given angular aperture is related to the galaxy count $N_g$ in that aperture via
\begin{equation}
    \label{eq:delta_g}
    \delta_g = \frac{N_g}{\bar{N}_g} - 1\,.
\end{equation}
Instead of the PDF $p(N_g, \kappa)$ we can hence equivalently study the PDF $p(\delta_g, \kappa)$. For a Eulerian bias model, we can directly parametrise the conditional mean in a Taylor expansion. To second order we obtain
\begin{equation}
    \label{eq:bias_model_Eulerian}
    \langle\delta_g|\delta_m\rangle = b_1^E \delta_m + \frac{b_2^E}{2} (\delta_m^2-\sigma_m^2)\,,
\end{equation}
where subtracting the matter variance term ensures a zero mean for the galaxy density contrast. In our analysis, we adopt this quadratic Eulerian bias expansion, which already predicts the measured conditional mean $\langle\delta_g|\delta_m\rangle$ with sub-percent level accuracy. To obtain the parameters, we fit the conditional mean of galaxy counts given dark matter density  along with the measured lens number density $n_g^{\rm 3D}=1.2 \times 10^{-3}$ Mpc$^{-3}$ corresponding to an angular density of $n_g=0.0834$ arcmin$^{-2}$. This leads to the parameters summarised in Table~\ref{tab:galaxy_bias_stochasticity}. Our fitted bias parameters are similar to the theoretical expectations for DES-like mock galaxy sample $b_1=1.93$, $b_{\delta^2}=0.19$ and $b_{s^2}=-0.53$ quoted in \cite{Halder_2023}. It is expected that the counts-in-cells $b_1$ parameter is scale-dependent and approaches the standard linear bias in the limit of increasing scale \citep{Friedrich2020}. The $b_{\delta^2}$ corresponds to our  $b_2/2$, but for circularly averaged densities the tidal bias term $s^2$ will correlate with the $\delta^2$ term thus lowering the effective $b_2$.

Lagrangian bias models can be implemented through the joint PDF and CGF of $\delta_m$ and $\delta_g$ (and suppressing all dependence on the smoothing aperture for now) for which this expectation value is given by
\begin{align}
\label{eq:expec_from_CGF}
    &\ \langle \delta_g | \delta_m \rangle \nonumber \\
    =&\ \frac{1}{p(\delta_m)} \int \dd \delta_g\ \delta_g\ p(\delta_g , \delta_m) \nonumber \\
    =&\ \frac{1}{p(\delta_m)} \int \frac{\dd\lambda_g\dd\lambda_m}{(2\pi)^2}\ e^{ - i\lambda_m\delta_m + \varphi(i\lambda_m, i\lambda_g)} \int \dd \delta_g\ \delta_g\ e^{-i\lambda_g\delta_g} \nonumber \\
    =&\ \frac{1}{p(\delta_m)} \int \frac{\dd\lambda_g\dd\lambda_m}{2\pi}\ e^{ - i\lambda_m\delta_m + \varphi(i\lambda_m, i\lambda_g)}\ i\frac{\dd \delta_{\mathrm D}(\lambda_g)}{\dd \lambda_g} \nonumber \\
    =&\ \frac{\int \frac{\dd\lambda_m}{2\pi}\ e^{ - i\lambda_m\delta_m + \varphi(i\lambda_m)}\ \partial_{\lambda_g}\varphi(i\lambda_m, 0)}{\int \frac{\dd\lambda_m}{2\pi}\ e^{ - i\lambda_m\delta_m + \varphi(i\lambda_m)}} \ .
\end{align}
Now for any given angular smoothing radius $\theta$, the projected CGF of galaxies and matter is given by a line-of-sight integral analogous to \eqnrefnospace{CGF_Limber}.
\begin{align}
    &\ \varphi_\theta(\lambda_m, \lambda_g)\nonumber \\ \approx &\ \int \dd \chi\ \lim_{L\rightarrow \infty} \frac{\varphi_{\mathrm{cy}, \chi\theta, L}\left(L w_m(\chi)\lambda_m, L w_{m}(\chi) \lambda_g, \eta_0 -\chi/c\right)}{L}\ ,
\end{align}
where $\varphi_{\mathrm{cy}, \chi\theta, L}(\cdot , \cdot, \eta)$ is now the joint CGF of matter and galaxy density contrast in cylinders of radius $\chi\theta$ and length $L$ at time $\eta$. To compute the conditional mean $\langle \delta_g|\delta_m\rangle$ from the conditional CDF \eqnref{expec_from_CGF} we only need its first derivative at the origin
\begin{align}
\label{eq:Limber_for_galaxies}
    &\ \left.\frac{\partial\varphi_\theta(\lambda_m, \lambda_g)}{\partial \lambda_g}\right|_{\lambda_g=0}\nonumber \\ \approx &\ \int \dd \chi\ w_m(\chi)\ \lim_{L\rightarrow \infty} \partial_g\varphi_{\mathrm{cy}, \chi\theta, L}\left(L w_m(\chi)\lambda_m,0, \eta_0 - \chi/c\right)\ ,
\end{align}
where $\partial_g\varphi_{\mathrm{cy}, \chi\theta, L}$ is the partial derivative of $\varphi_{\mathrm{cy}, \chi\theta, L}$ \wrt its second argument. This derivative can be calculated along the lines of \citet{Friedrich2021b}. These results described in Section~\ref{sec:cylinder_CGF} can be summarized as follows: Compute the minimum of the action of the functional integral featuring linear and nonlinear densities
\begin{equation}
    \lbrace \delta^*, j^* \rbrace = \underset{\delta, j}{\mathrm{argmin}}\ s_{\lambda_m}(\delta, j)\ ,
\end{equation}
where $s_\lambda$ will be defined in \eqnrefnospace{2D_function_to_be_minimized}. 
Then any given Lagrangian-style bias model $f_L(\delta^*)$ for the derivative of the CGF which we need in \eqnref{Limber_for_galaxies} is given by
\begin{equation}
    \partial_g\varphi_{\mathrm{cy}, \chi\theta, L}(\lambda_m, 0) \approx \left(1+\mathcal{F}(\delta^*)\right) f_L(\delta^*) - 1\ .
\end{equation}
\citet{Friedrich2021b} have demonstrated that the quadratic function
\begin{equation}
    f_L(\delta^*)=1 + b_1^L \delta^* + \frac{b_2^L}{2} (\delta^*)^2\,, 
\end{equation}
can be thought of as a second order Lagrangian bias expansion for $\partial_g\varphi_{\mathrm{cy}, \chi\theta, L}(\lambda_m, 0)$ and hence for $\langle \delta_g | \delta_m \rangle$. The coefficients $b_1^L$ and $b_2^L$ are then the linear and quadratic Lagrangian bias parameters. While the zero mean for $\langle \delta_g | \delta_m \rangle$ has to be ensured manually in the Eulerian model, it is automatic in the CGF evaluation since the Lagrangian-to-Eulerian mapping  is built into our path integral formulation. As expected the fitted values of the quadratic Eulerian and Lagrangian bias parameters are roughly related by $b_1^E\approx 1+b_1^L$ and $b_2^E\approx \frac{\nu-1}{\nu}b_1^L+b_2^L$ with the collapse index $\nu=21/13$.

Following results for spectroscopic tracers presented in \cite{Gould2025}, it might be beneficial to adopt a Gaussian Lagrangian bias description \citep[first described in][]{Stuecker2024_cumulantbias,Stuecker2024}
\begin{equation}
    \label{eq:bias_L_Gauss}
f_{L}(\delta^*)=\frac{\exp\left[-\frac{(b_1^{\rm G})^2}{2b_2^{\rm G}}\right]}{\sqrt{1+b_2^{\rm G}\sigma_m^2}} \exp\left[\frac{b_2^{\rm G}\left(\frac{b_1^{\rm G}}{b_2^{\rm G}}+\delta^*\right)^2}{2(1+b_2^{\rm G}\sigma_m^2)}\right]\,,
\end{equation}
where the renormalised Gaussian Lagrangian bias parameters $b_n^{\rm G}$ are expected to be close to scale-independent. We find that $b_1^L\approx \tilde b_1^G$ and $b_2^L\approx (\tilde b_1^G)^2+ \tilde b_2^G$, where $\tilde b_n^G=b_n^G/(1+b_2^G\sigma_m^2)$ are the scale-dependent parameters.

Figure~\ref{fig:Bias_residual} compares the residuals between the fitted and measured $\langle \delta_g|\delta_m\rangle$, and shows that the quadratic Eulerian and Lagrangian bias expansions yield similar results at the 0.2\% level, while the Gaussian Lagrangian bias model  provides further improved accuracy.

\begin{table}
    \centering
    \begin{tabular}{c|c|c|c|c|c}
       $b_1^E$  &  $b_2^E$ & $b_1^L$  & $b_2^L$ & $b_1^G$  & $b_2^G$ \\\hline 
        2.045  & 0.230 & 1.048 & -0.131 & 1.034  & -1.261 
    \end{tabular}
        \begin{tabular}{c|c|c}
       $\alpha_0$ & $\alpha_1$ & $\alpha_2$ \\\hline 
 1.453 & -0.104 & 1.720 
    \end{tabular}
    \caption{Fitted quadratic galaxy bias parameters for the Eulerian, Lagrangian and Gaussian Lagrangian models (top) and quadratic stochasticity parameters (bottom) for the lens sample.}
    \label{tab:galaxy_bias_stochasticity}
\end{table}

\subsection{Galaxy stochasticity (shot-noise)}
\label{sec:shot-noise}
Following the previous subsection, this means that we have to calculate \eqref{eq:jointPDF}
where the conditional PDF $p(N_g|\delta_m)$ encodes the shot-noise or stochasticity of galaxies around the expectation value \eqref{eq:bias}.

It is often assumed that this shot-noise is Poissonian \citep[\eg][]{Efstathiou1995, Clerkin2017, Salvador2019, Repp2020}. However, both in simulated and in observational data, it has been found that this assumption is not accurate \citep[\egnospace][]{Hamaus2011, Friedrich2018, Gruen2018, Friedrich2021b}. So we instead adopt the approach of \citet{Friedrich2018} and \citet{Gruen2018}, who proposed the following generalisation of the Poisson distribution:
\begin{equation}
    P_\alpha(N_g|\delta_m) = \mathcal{N} \exp\left\lbrace \frac{N_g}{\alpha}\ln\left[\frac{\bar N_g}{\alpha}\right] - \ln \Gamma \left[\frac{N_g}{\alpha}+1\right] - \frac{\bar N_g}{\alpha}\right\rbrace\,,
\end{equation}
where we use shorthand notation $N_g=N_g(\delta_m)=\bar N_g(1+\langle \delta_g|\delta_m\rangle)$ and $\alpha=\alpha(\delta_m)$.
Here $\mathcal{N}$ is a normalisation constant that is very closely approximated by $1/\alpha$, and $\alpha$ parametrizes deviation from the Poisson distribution with $\alpha=1$ representing the Poisson case. In \citet{Friedrich2018, Friedrich2021b,Britt2024} it was shown that $\alpha$ needs to be a function of the underlying matter density contrast in order to accurately capture the shot-noise behaviour of simulated galaxies. Here, we are using a quadratic ansatz 
\begin{equation}
\label{eq:stochasticity_model}
\frac{\langle N_g^2|\delta_m\rangle_c}{\langle N_g |\delta_m\rangle}=\alpha(\delta_m) = \alpha_0 + \alpha_1\delta_m +\alpha_2\delta_m^2\,,
\end{equation}
which proved effective for halos and HOD mock galaxies in both photometric and spectroscopic settings \citep{Friedrich2021b,Britt2024,Gould2025}. Fitting the corresponding parameters leads to an agreement at the percent level, as shown in the lower panel of Figure~\ref{fig:Bias_residual} with the values stated in Table~\ref{tab:galaxy_bias_stochasticity}. 

\subsection{Galaxy shape noise}\label{sec:shape_noise}
Galaxy surveys do not provide direct access to the weak lensing convergence $\kappa$, but instead the shear components of individual galaxies. As galaxies have intrinsic shapes, the observed weak lensing signal is a combination of the cosmic shear and the intrinsic galaxy ellipticity. Assuming that the intrinsic ellipticity is independent of the cosmic shear signal, the total weak lensing convergence is a sum of two independent variables and its PDF is hence a convolution of the individual distributions. 
\begin{equation}
\label{eq:pdf_noisy}
    p_{\rm{noisy} }(\kappa) = (p_{\rm SN}\star p)(\kappa)\,.
\end{equation}
This property extends straightforwardly to the joint PDF $p(N_g,\kappa)$, as the galaxy count $N_g$ is unaffected by shape noise.
The amount of noise coming from the intrinsic shapes is controlled by the source number density and the size of the aperture. Often, the shape noise can be regarded as a Gaussian of variance $\sigma_{\rm SN}^2$
\begin{equation}
    \label{eq:shapeNoise_PDF}
   p_{\rm SN}(\kappa_{\rm SN}) = \frac{1}{\sqrt{2 \pi }} \exp\left(-\frac{\kappa_{\rm SN}^2}{2\sigma_{\rm SN}^2}\right) \,,\ \sigma_{\rm SN}^2 = \frac{\sigma_{\epsilon}^2}{n_g \Omega_{\theta}} \,,
\end{equation}
with $n_g$ the source galaxy number density, $\sigma_\epsilon^2$ the intrinsic ellipticity noise and $\Omega_{\theta}$ the solid angle. For Gaussian shape noise, its effect can be incorporated by adding the shape noise variance to the intrinsic variance of $\kappa$ in the joint CGF of $\delta_m$ and $\kappa$, which is the approach taken here. By performing random rotations of galaxy shapes, the shape noise can be isolated and its variance (or full one-point PDF $p_{\rm SN}$) entering in equation~\eqref{eq:pdf_noisy} can be measured. The inclusion of further weak lensing systematics relating to the convergence reconstruction from a masked shear field, multiplicative shear bias, baryonic effects, and intrinsic alignments is discussed in \cite{Barthelemy2024} and will be validated against simulations in upcoming work.

This completes our model for the joint PDF $p(N_g, \kappa)$ and we now turn to practical considerations for evaluating this model. In the next two subsections we detail the formulas implemented in the {\verb|CosMomentum|} code. Readers may feel free to skip those and proceed with the model validation in Section~\eqref{sec:validation}.

\subsection{The CGF of density contrast in long cylinders}
\label{sec:cylinder_CGF}

To simplify the notation, we will suppress the time dependence of all quantities in this subsection. \Eg instead of $\varphi_{\mathrm{cy}, R, L}(\lambda, \eta)$ we will only write $\varphi_{\mathrm{cy}, R, L}(\lambda)$. The time dependence of the final result will be evident from the derivation and we will explicitly highlight it at the end of the subsection.

\subsubsection{The reduced CGF in the quasi-linear limit}

Let $\delta_{R,L}$ be the 3-dimensional matter density contrast averaged over a cylindrical aperture of radius $R$ and length $L$. The cumulant generating function $\varphi_{\mathrm{cy}, R, L}(\lambda)$ of $\delta_{R,L}$ is then defined through
\begin{equation}
    \varphi_{\mathrm{cy}, R, L}(\lambda) = \sum_{n\geq 2} \frac{\langle \delta_{R,L}^n \rangle_c}{n!}\ \lambda^n\ .
\end{equation}
To devise a theoretical model for this CGF, it is useful to consider it as being constructed from two quantities: (i) the variance of density fluctuations $\sigma_{R,L}^2 \equiv \langle \delta_{R,L}^2 \rangle_c$ and, (ii) the so-called reduced cumulant generating function
\begin{equation}
\label{eq:reduced_CGF_cyl}
    \tilde\varphi_{\mathrm{cy}, R, L}(\lambda) \equiv \sum_{n\geq 2} \frac{S_n}{n!}\ \lambda^n\ ,
\end{equation}
where
\begin{equation}
    S_n \equiv  \frac{\langle \delta_{R,L}^n \rangle_c}{\sigma_{R,L}^{2(n-1)}}\,,
\end{equation}
are the so-called reduced cumulants. The unreduced CGF is then related to the two ingredients as
\begin{equation}
\label{eq:reduced_to_unreduced_CGF}
    \varphi_{\mathrm{cy}, R, L}(\lambda) = \tilde\varphi_{\mathrm{cy}, R, L}(\sigma_{R,L}^2\lambda)/\sigma_{R,L}^2\ .
\end{equation}
The reduced cumulants, and correspondingly the reduced CGF $\tilde\varphi_{\mathrm{cy}, R, L}(\lambda)$ have a well-defined, non-zero limit $\sigma_{R,L}^2\rightarrow 0$. We follow \citet{Valageas2002} and earlier work \eg by \cite{Bernardeau1994} in calling this the quasi-linear limit and we are computing this limit in the remainder of this subsection. To calculate $\sigma_{R, L}^2$ in \eqnrefnospace{reduced_to_unreduced_CGF}, we will be using the revised \emph{halofit} prescription of \citet{Takahashi2012} for the non-linear power spectrum \citep[see][for the original halofit]{Smith2003}.

\subsubsection{Functional integral for the reduced CGF}

The reduced CGF defined in \eqnref{reduced_CGF_cyl} can also be expressed through the following expectation value (\cf \eqnrefnospace{CGF_as_expectation}),
\begin{align}
e^{\tilde\varphi_{\mathrm{cy}, R, L}(\sigma_{R, L}^2\lambda)/\sigma_{R, L}^2} =&\ \langle e^{\lambda \delta_{R,L}} \rangle\nonumber \\
\Rightarrow\ \ \ \ \ \  e^{\tilde\varphi_{\mathrm{cy}, R, L}(\lambda)/\sigma_{R, L}^2} =&\ \langle e^{\lambda \delta_{R,L} / \sigma_{R, L}^2} \rangle\ .
\end{align}
At any given point in space, $\delta_{R,L}$ is uniquely determined by the configuration of the initial cosmic density field, or equivalently: by today's linear density contrast $\delta_{\mathrm{lin}}(\bm{x})$. Hence, in the spirit of \citet{Valageas2002} and following the notation shown in \citet{Friedrich2020}, who performed similar calculations for a spherical aperture, we can write the above expectation value as a functional integral over all  possible configurations of $\delta_{\mathrm{lin}}$,
\begin{align}
    e^{\tilde\varphi_{\mathrm{cy}, R, L}(\lambda)/\sigma_{R, L}^2} =\ & \int \mathcal{D}\delta_{\mathrm{lin}}\ \mathcal{P}[\delta_{\mathrm{lin}}]\ \exp\left(\frac{\lambda \delta_{R,L}[\delta_{\mathrm{lin}}]}{\sigma_{R, L}^2}\right)\ .
\end{align}
Here $\mathcal{P}[\delta_{\mathrm{lin}}]$ is the probability distribution functional of the linear density contrast field and the non-linear density contrast $\delta_{R,L}$ has been expressed as a functional of $\delta_{\mathrm{lin}}$. Note again that we dropped the time dependence from our notation - this dependence is solely carried by the functional $\delta_{R,L}[\cdot] = \delta_{R,L}[\cdot, \eta]$ since $\delta_{\mathrm{lin}}$ is always considered at $\eta_0$.

For Gaussian initial conditions, the probability density functional $\mathcal{P}[\delta_{\mathrm{lin}}]$ is a Gaussian functional that is completely determined by the linear power spectrum $P_{\mathrm{lin}}(k)$ \citep[\cfnospace][]{Valageas2002}. For general non-Gaussian initial conditions we instead follow \citet{Friedrich2020} and express $\mathcal{P}[\delta_{\mathrm{lin}}]$ through the linear cumulant generating functional 
\begin{equation}
    \Phi[J_{\mathrm{lin}}] = \sum_{n=1}^\infty\ \frac{1}{n!} \int \prod_{i=1}^n\ \dd^3 x_i\ J_{\mathrm{lin}}(\mathbf{x}_i)\ \xi_{{\mathrm{lin}},n}(\mathbf{x}_1,\ \dots\ , \mathbf{x}_n)\ .
\end{equation}
Here $\xi_{{\mathrm{lin}},n}$ are the connected $n$-point correlation functions of the linear density field \citep[see \egnospace][for a definition of connected n-point correlators]{BernardeauReview}. Using the fact that $\mathcal{P}$ and $\Phi$ are related through a functional Laplace transform, we can now express $\tilde\varphi_{\mathrm{cy}, R, L}$ as
\begin{align}
\label{eq:Friedrich_path_integral}
    & e^{\tilde\varphi_{\mathrm{cy}, R, L}(\lambda)/\sigma_{R, L}^2} = \nonumber \\
    & \frac{1}{\mathcal{N}} \int \mathcal{D}\delta_{\mathrm{lin}}\ \mathcal{D}J_{\mathrm{lin}}\ \exp\left(\frac{\lambda \delta_{R,L}[\delta_{\mathrm{lin}}]}{\sigma_{R, L}^2} - iJ_{\mathrm{lin}} \cdot \delta_{\mathrm{lin}} + \Phi[iJ_{\mathrm{lin}}]\right)\ .
\end{align}
The linear cumulant generating functional $\Phi[J_{\mathrm{lin}}]$ encodes the initial conditions of the density field. For Gaussian initial fluctuations, it is solely determined by the linear two-point correlation function. In the present paper, we restrict ourselves to this Gaussian case, but we nevertheless continue with the notation of \citet{Friedrich2020} because it streamlines some of the derivations and because we intend to investigate primordial non-Gaussianity in a follow-up paper.

The normalisation constant $\mathcal{N}$ on the right-hand side of \eqnref{Friedrich_path_integral} is given by $|2\pi\mathds{1}_J|$, \ie the determinant of $2\pi$ times the unit operator in the space of $J_{\mathrm{lin}}$. This is formally infinite but as for \citet{Friedrich2020}, the normalisation will eventually drop in our calculations. Note furthermore that we have used the abbreviation
\begin{equation}
    J_{\mathrm{lin}} \cdot \delta_{\mathrm{lin}} \equiv \int \dd^3 x\ J_{\mathrm{lin}}(\mathbf{x})\ \delta_{\mathrm{lin}}(\mathbf{x})\ .
\end{equation}

\subsubsection{Identifying a large-deviation parameter}

We would like to evaluate \eqnref{Friedrich_path_integral} by means of a saddle-point approximation (also known as Laplace's method). This means that we would like to re-express the equation as
\begin{align}
\label{eq:Laplace_path_integral}
    & e^{\tilde\varphi_{\mathrm{cy}, R, L}(\lambda)/\sigma_{R, L}^2} \propto\nonumber \\
    &\ \ \ \ \ \ \ \ \ \int \mathcal{D}\delta_{\mathrm{lin}}\ \mathcal{D} \tilde J_{\mathrm{lin}}\ \exp\left\lbrace-\frac{1}{x} \tilde S_\lambda[\delta_{\mathrm{lin}}, \tilde J_{\mathrm{lin}}]\right\rbrace\ ,
\end{align}
where we have anticipated a re-definition $J_{\mathrm{lin}} \rightarrow \tilde J_{\mathrm{lin}}$ and where $x$ is a small parameter such that the action $\tilde S_\lambda[\delta_{\mathrm{lin}}, \tilde J_{\mathrm{lin}}]$ becomes independent of $x$ in the limit $x\rightarrow 0$. We call $x$ the large-deviation or driving parameter \citep{Bernardeau_2016} and for small $x$, a good approximation of the reduced CGF would be given by
\begin{equation}
    \tilde\varphi_{\mathrm{cy}, R, L}(\lambda) \approx -\frac{\sigma_{R, L}^2}{x}\ \tilde S_\lambda[\delta_{\mathrm{lin}}^*, \tilde J_{\mathrm{lin}}^*]\ , 
\end{equation}
where $\delta_{\mathrm{lin}}^*$ and $\tilde J_{\mathrm{lin}}^*$ are the configurations of the fields $\delta_{\mathrm{lin}}$ and $\tilde J_{\mathrm{lin}}$ that minimize $\tilde S_\lambda$ (the saddle-point configurations).

As a first step, let us rewrite \eqnref{Friedrich_path_integral} as
\begin{align}
    e^{\tilde\varphi_{\mathrm{cy}, R, L}(\lambda)/\sigma_{R, L}^2} =\ & \frac{1}{\mathcal{N}} \int \mathcal{D}\delta_{\mathrm{lin}}\ \mathcal{D}J_{\mathrm{lin}}\ e^{-S_\lambda[\delta_{\mathrm{lin}}, J_{\mathrm{lin}}]}\,,
\end{align}
with the intermediate action
\begin{equation}
\label{eq:def_of_action}
    S_\lambda[\delta_{\mathrm{lin}}, J_{\mathrm{lin}}] \equiv - \frac{\lambda \delta_{R,L}[\delta_{\mathrm{lin}}]}{\sigma_{R, L}^2} + iJ_{\mathrm{lin}} \cdot \delta_{\mathrm{lin}} - \Phi[iJ_{\mathrm{lin}}]\ .
\end{equation}
An obvious candidate for the large-deviation parameter is $x = \sigma_{R, L}^2$. But the last term on the right-hand side of \eqnref{def_of_action} does not become independent of $\sigma_{R, L}^2$ in the limit of $\sigma_{R, L}^2 \rightarrow 0$. In that limit, $\sigma_{R, L}^2$ is approximately proportional to $\sigma_8^2$ (standard deviation of linear density contrast in spheres of $8$Mpc$/h$, which is commonly used to parametrize the amplitute of the linear power spectrum) and the $n$th order correlation functions appearing in the definition of $\Phi[J_{\mathrm{lin}}]$ are proportional to $\sigma_8^{2(n-1)}$ \citep[\cfnospace][and references therein]{BernardeauReview}. Given any particular configuration of $J_{\mathrm{lin}}$, a quantity that is independent of $\sigma_8$ (again, in the limit of $\sigma^2\rightarrow 0$) is
\begin{equation}
   \tilde \Phi[J_{\mathrm{lin}}] \equiv \sigma_{R, L}^2\Phi[ J_{\mathrm{lin}}/\sigma_{R, L}^2] \ .
\end{equation}
Using this definition, $S_\lambda$ turns into
\begin{align}
     S_\lambda[\delta_{\mathrm{lin}}, J_{\mathrm{lin}}] =&\ - \frac{\lambda \delta_{R,L}[\delta_{\mathrm{lin}}]}{\sigma_{R, L}^2} + iJ_{\mathrm{lin}} \cdot \delta_{\mathrm{lin}} - \frac{\tilde \Phi[i\sigma_{R, L}^2 J_{\mathrm{lin}}]}{\sigma_{R, L}^2}\nonumber \\
     \equiv&\ \frac{1}{\sigma_{R, L}^2} \left[- \lambda \delta_{R,L}[\delta_{\mathrm{lin}}] + i\tilde J_{\mathrm{lin}} \cdot \delta_{\mathrm{lin}} - \tilde \Phi[i\tilde J_{\mathrm{lin}}]\right]\ .
\end{align}
\eqnref{Friedrich_path_integral} hence becomes
\begin{align}
\label{eq:Laplace_path_integral_final}
    & e^{\tilde\varphi_{\mathrm{cy}, R, L}(\lambda)/\sigma_{R,L}^2} =\nonumber \\
    & \ \ \ \ \ \ \frac{1}{\widetilde{\mathcal{N}}}\int \mathcal{D}\delta_{\mathrm{lin}}\ \mathcal{D} \tilde J_{\mathrm{lin}}\ \exp\left\lbrace-\frac{1}{\sigma_{R,L}^2} \tilde S_{\lambda}[\delta_{\mathrm{lin}}, \tilde J_{\mathrm{lin}}]\right\rbrace\,,
\end{align}
with a new action
\begin{equation}
\label{eq:new_action}
    \tilde S_{\lambda}[\delta_{\mathrm{lin}}, \tilde J_{\mathrm{lin}}] \equiv - \lambda \delta_{R,L}[\delta_{\mathrm{lin}}] + i\tilde J_{\mathrm{lin}} \cdot \delta_{\mathrm{lin}} - \tilde \Phi[i\tilde J_{\mathrm{lin}}]\,,
\end{equation}
and a new normalisation constant $\widetilde{\mathcal{N}} = |2\pi\sigma_{R,L}^2\mathds{1}_J|$, which is again formally infinite. Now we are in a position to apply Laplace's method and derive the approximation
\begin{align}
    e^{\tilde\varphi_{\mathrm{cy}, R, L}(\lambda)/\sigma_{R,L}^2} \approx\ & \frac{1}{\mathcal{A}^{1/2}}\  \exp\left\lbrace-\tilde S_\lambda[\delta_{\mathrm{lin}}^*, \tilde J_{\mathrm{lin}}^*]/\sigma_{R,L}^2\right\rbrace\ .
\end{align}
Here $\mathcal{A}$ is the determinant of the Hessian matrix of the functional $\tilde S_\lambda$ (when considered as a matrix in the combined space of $\delta_{\mathrm{lin}}$ and $\tilde J_{\mathrm{lin}}$) evaluated at the saddle-point configurations $\delta_{\mathrm{lin}}^*$ and $\tilde J_{\mathrm{lin}}^*$ that minimize the action $\tilde S_\lambda$. Note that the normalisation $\widetilde{\mathcal{N}}$ has now been cancelled by what is formally an infinite number of factors of $2\pi\sigma_{R,L}^2$ resulting from the infinite dimensional saddle-point approximation. The reduced CGF of matter density contrast averaged in the cylindrical aperture is then given by
\begin{equation}
\label{eq:saddle_point_approximation_for_CGF}
    \tilde\varphi_{\mathrm{cy}, R, L}(\lambda) \approx  -\tilde S_\lambda[\delta_{\mathrm{lin}}^*, \tilde J_{\mathrm{lin}}^*] - \frac{\sigma_{R,L}^2}{2} \ln \mathcal{A}[\delta_{\mathrm{lin}}^*, \tilde J_{\mathrm{lin}}^*]\ .
\end{equation}

There are two ways to use this approximation. Restricting to the first term on the right-hand side of \eqnref{saddle_point_approximation_for_CGF} and computing $\sigma_{R,L}^2$ at the leading-order (tree-level) in perturbation theory in the definitions of $\tilde S_\lambda$ and $\tilde \Phi$ yields the corresponding reduced cumulant generating function at tree-level in perturbation theory \citep{Valageas2002, Valageas2002V}. This means that the resulting reduced CGF will be equivalent to
\begin{equation}
    \tilde\varphi_{\mathrm{cy}, R, L}(\lambda) \equiv \sum_{n\geq 2} \frac{S_n^{\mathrm{tree}}}{n!}\ \lambda^n\ ,
\end{equation}
where all reduced cumulants are evaluated at tree-level in standard perturbation theory. This is the approach we take, and we demonstrate in \secref{validation} that it yields accurate predictions for the PDF $p(\delta_g, \kappa)$ on the scales we consider.

Another approach would be to use both terms in \eqnref{saddle_point_approximation_for_CGF} and to compute $\sigma_{R,L}^2$ at the next-to-leading order in perturbation theory (the 1-loop order) in the definitions of $\tilde S_\lambda$ and $\tilde \Phi$. This would yield the reduced cumulant generating function at the 1-loop order in standard perturbation theory \citep{Valageas2002V}. That would require us to evaluate the functional determinant $\mathcal{A}[\delta_{\mathrm{lin}}^*, \tilde J_{\mathrm{lin}}^*]$ which is both numerically and analytically expensive. \citet{Ivanov2019} have demonstrated how to evaluate a determinant analogous to $\mathcal{A}$ in a direct calculation of the PDF of matter density fluctuations in 3D spheres. Unfortunately, their results do not directly carry over to our situation, since we need the CGF (and we need it in cylindrical filters) in order to perform the line-of-sight projection of \eqnrefnospace{CGF_Limber}. 

\subsubsection{Minimizing the action $\tilde S_\lambda$}
\label{sec:minimizing_the_action}

We continue to follow the notation of \citet{Friedrich2020} and denote functional derivation of a functional $F$ \wrt the function $f$ as $\mathpzc{d}F/\mathpzc{d}f(\mathbf{x})$. To minimize the action defined in \eqnref{new_action} we need to find configurations $\delta_{\mathrm{lin}}^*$ and $\tilde J_{\mathrm{lin}}^*$ such that
\begin{equation}
    \left.\frac{\mathpzc{d} \tilde S_\lambda}{\mathpzc{d} \delta_{\mathrm{lin}}(\mathbf{x})}\right|_{\delta_{\mathrm{lin}}^*, \tilde J_{\mathrm{lin}}^*} =\ \ 0\ \ = \left.\frac{\mathpzc{d} \tilde S_\lambda}{\mathpzc{d} \tilde J_{\mathrm{lin}}(\mathbf{x})}\right|_{\delta_{\mathrm{lin}}^*, \tilde J_{\mathrm{lin}}^*}\\
\end{equation}
\begin{align}
\label{eq:minimisation_A}
    \Rightarrow\ i\tilde J_{\mathrm{lin}}^*(\mathbf{x}) =\ & \lambda\left.\frac{\mathpzc{d} \delta_{R,L}}{\mathpzc{d} \delta_{\mathrm{lin}}(\mathbf{x})}\right|_{\delta_{\mathrm{lin}}^*}\,,\\
\label{eq:minimisation_B}
    \delta_{\mathrm{lin}}^*(\mathbf{x}) =\ & \left.\frac{\mathpzc{d} \tilde \Phi}{\mathpzc{d} \tilde J_{\mathrm{lin}}(\mathbf{x})}\right|_{i\tilde J_{\mathrm{lin}}^*}\ .
\end{align}
For long cylinders (\ie $L \gg R$, which is the relevant limit in \eqnrefnospace{CGF_Limber}) one can argue in analogy to \citet{Valageas2002} and \citet{Friedrich2020} that the saddle-point configurations $\delta_{\mathrm{lin}}^*$ and $\tilde J_{\mathrm{lin}}^*$ will be cylindrically symmetric. On the space of cylindrically symmetric configurations the functional $\delta_{R,L}[\delta_{\mathrm{lin}}]$ is completely determined through cylindrical collapse. Especially we have
\begin{align}
\label{eq:delta_R_equals_F}
    \delta_{R,L}[\delta_{\mathrm{lin}}] =\ & \mathcal{F}(\delta_{{\mathrm{lin}},R_{\mathrm{lin}}})\ ,
\end{align}
where the function $\mathcal{F}(\delta_{{\mathrm{lin}},R_{\mathrm{lin}}})$ describes the cylindrically collapsed density fluctuation at radius $R$ that corresponds to a linear density fluctuation $\delta_{{\mathrm{lin}},R_{\mathrm{lin}}}$ at the initial, linear radius
\begin{align}
\label{eq:linear_radius}
     R_{\mathrm{lin}} =\ & R\ (1+\delta_{R,L}[\delta_{\mathrm{lin}}])^{1/2}\ .
\end{align}
The differential equation determining this function $\mathcal{F}$ is given in \apprefnospace{spherical_collapse}. Note that so far the only changes \wrt \citet{Friedrich2020} are the fact that $1+\delta_{R,L}$ appears with an exponent of $1/2$ in \eqnref{linear_radius} (as opposed to $1/3$ for the spherical aperture), that $\mathcal{F}$ describes cylindrical collapse (as opposed to spherical collapse) and that in the previous subsection we specifically factored out the large-deviation parameter $\sigma_{R,L}^2$. Hence, by following the derivations of \citet{Friedrich2020} together with these modifications, one arrives at
\begin{align}
 \lambda\left.\frac{\mathpzc{d} \delta_{R,L}}{\mathpzc{d} \delta_{\mathrm{lin}}(\mathbf{x})}\right|_{\delta_{\mathrm{lin}}^*} =\ & \lambda\frac{\mathcal{F}'(\delta_{{\mathrm{lin}},R_{\mathrm{lin}}}^*)\ W_{R_{\mathrm{lin}},L} (\mathbf{x})}{1 - \mathcal{F}'(\delta_{{\mathrm{lin}},R_{\mathrm{lin}}}^*) \left.\frac{\dd \delta_{{\mathrm{lin}},R'}^*}{\dd R'}\right|_{R_{\mathrm{lin}}} \frac{R^2}{2R_{\mathrm{lin}}}} \nonumber \\
 =:\ & A_\lambda[\delta_{\mathrm{lin}}^*]\ W_{R_{\mathrm{lin}},L} (\mathbf{x})\\
 \label{eq:solution_for_J_star}
 \Rightarrow\ i\tilde J_{\mathrm{lin}}^*(\mathbf{x}) =\ & A_\lambda[\delta_{\mathrm{lin}}^*]\ W_{R_{\mathrm{lin}},L} (\mathbf{x})\ ,
\end{align}
where $W_{R,L} (\mathbf{x})$ is the top-hat filter corresponding to averaging over cylinders of radius $R$ and length $L$. Using \eqnref{solution_for_J_star} one can then show \citep[again in complete analogy to the derivations of][for the spherical case]{Friedrich2020} that the action of the functional integral at its minimum is given by
\begin{multline}
    \tilde S_\lambda[\delta_{\mathrm{lin}}^*, \tilde J_{\mathrm{lin}}^*] =\ -\lambda \mathcal{F}(\delta_{{\mathrm{lin}},R_{\mathrm{lin}}}^*) + A_\lambda\ \delta_{{\mathrm{lin}},R_{\mathrm{lin}}}^*  \\
    \ - \sigma_{R,L}^2\ \varphi_{\mathrm{lin}, R_{\mathrm{lin}}, L}\left(\frac{A_\lambda}{\sigma_{R,L}^2}\right)\ ,
\end{multline}    
where $\varphi_{\mathrm{lin}, R_{\mathrm{lin}}, L}$ is the cumulant generating function of the linear density field, when averaged in cylinders of radius $R_{\mathrm{lin}}$ and length $L$. As for the spherical case, this is in fact equal to the minimum of the 2-dimensional function
\begin{align}
\label{eq:2D_function_to_be_minimized}
    s_\lambda(\delta, j) =&\ -\lambda \mathcal{F}(\delta) + j \delta \nonumber \\
    &\ - \sigma_{R,L}^2\ \varphi_{{\mathrm{lin}}, R(1+\mathcal{F}(\delta))^{1/2}, L}\left(\frac{j}{\sigma_{R,L}^2}\right)\ ,
\end{align}
since the location of this minimum is given by the equations
\begin{align}
    j^* =\ & \mathcal{F}'(\delta^*)\left( \lambda  + \sigma_{R,L}^2\left.\frac{\dd\varphi_{\mathrm{lin}, R', L}}{\dd R'}\left(\frac{j^*}{\sigma_{R,L}^2}\right)\right|_{R'=R_\mathrm{lin}}\ \frac{R^2}{2R_\mathrm{lin}}\right)\,,\\
    \delta^* =\ & \sigma_{R,L}^2\left.\frac{\dd \varphi_{\mathrm{lin}, R_\mathrm{lin}, L}(j)}{\dd j}\right|_{j=(\nicefrac{j^*}{\sigma_{R,L}^2})}\,,
\end{align}
which have the solutions 
\begin{align}
    \delta^* = \delta_{\mathrm{lin},R_{\mathrm{lin}}}^*\ ,\ j^* = A_\lambda[\delta_{\mathrm{lin}}^*]\ .
\end{align}
Note that section 4.6 of \citet{Friedrich2020} spells out in detail how the above calculations can be performed in practice (modulo the differences between their and our situation mentioned before). We also give a brief outline of this in our \secrefnospace{practical_CGFevaluation}.

\subsection{Practical evaluation of the CGF and its line-of-sight projection}\label{sec:practical_CGFevaluation}

Let $\varphi_{\mathrm{cy}, R, L}(\lambda)$ again be the CGF of matter density contrast smoothed over a cylindrical aperture of radius $R$ and length $L$, \ie
\begin{equation}
    \varphi_{\mathrm{cy}, R, L}(\lambda) = \sum_{n=2}^\infty \langle (\delta_{\mathrm{cy}, R, L})^n \rangle_c\ \frac{\lambda^n}{n!}\ .
\end{equation}
To evaluate the Limber-type approximation of \eqnref{CGF_Limber}, we need to compute limits such as
\begin{equation}
    \lim_{L\rightarrow \infty} \frac{\varphi_{\mathrm{cy}, \chi\theta, L}\left(w(\chi)L\lambda, w\right)}{L}\ .
\end{equation}
This can be considered the equivalent of the Limber approximation \citep{Limber1953} in 2-point statistics for the CGF. To evaluate the above limit, let us recall how the cylindrical CGF is calculated. For long cylinders we consider the function
\begin{equation}
    s_\lambda(\delta , j) = -\lambda \mathcal{F}_{\mathrm{2D}}(\delta) + \delta j - \varphi_{\mathrm{lin}, R\sqrt{1+\mathcal{F}_{\mathrm{2D}}(\delta)}, L}(j)\ ,
\end{equation}
where \smash{$\varphi_{\mathrm{lin}, R\sqrt{1+\mathcal{F}_{\mathrm{2D}}(\delta)}, L}$} is the CGF of the linear density contrast averaged over cylinders of length $L$ and radius $R\sqrt{1+\mathcal{F}_{\mathrm{2D}}(\delta)}$ and where $\mathcal{F}_{\mathrm{2D}}(\delta)$ is the density contrast resulting from a linear density contrast $\delta$ through cylindrical collapse (\cf \apprefnospace{spherical_collapse}; also, note that we suppress time dependence here and in the following for more concise notation). We then approximate $\varphi_{\mathrm{cy}, R, L}$ by evaluating $-s_\lambda(\delta^*, j^*)$ , where $(\delta^*,j^*)$ is the minimum of $s_\lambda$. This amounts to solving the equations
\begin{align}
    \label{eq:numerical_tricks_1}
    j^* =&\ \lambda \mathcal{F}_{\mathrm{2D}}^{\ '}(\delta^*) + \left.\frac{\dd \varphi_{\mathrm{lin},R',L}(j^*)}{\dd R'}\right|_{R'=R_{\mathrm{lin}}}\ \frac{R^2}{2 R_{\mathrm{lin}}}\ \mathcal{F}_{\mathrm{2D}}^{\ '}(\delta^*)\,,\\
    \label{eq:numerical_tricks_2}
    \delta^* =&\ \left. \frac{\dd \varphi_{\mathrm{lin},R_{\mathrm{lin}}, L}(j)}{\dd j}\right|_{j=j^*}\,,\\
    R_{\mathrm{lin}} =&\ R\sqrt{1+\mathcal{F}_{\mathrm{2D}}(\delta)} \ .
\end{align}
In practice, it is easier to start with a given value of $\delta^*$ and then find the corresponding values of $j^*$ and $\lambda$. For small primordial non-Gaussianity we can approximate
\begin{equation}
    \varphi_{\mathrm{lin},R_{\mathrm{lin}}, L}(j) \approx \ \frac{\langle \delta_{\mathrm{lin},R_{\mathrm{lin}}, L}^2 \rangle}{2}\ j^2 + \frac{\langle \delta_{\mathrm{lin},R_{\mathrm{lin}}, L}^3 \rangle}{6}\ j^3 \,,
\end{equation}
\begin{equation}
    \frac{\dd \varphi_{\mathrm{lin},R_{\mathrm{lin}}, L}(j)}{\dd j} \approx \ \langle \delta_{\mathrm{lin},R_{\mathrm{lin}}, L}^2 \rangle\ j + \frac{\langle \delta_{\mathrm{lin},R_{\mathrm{lin}}, L}^3 \rangle}{2}\ j^2 \ .
\end{equation}
\begin{equation}
    \Rightarrow j^* \approx  \frac{\langle \delta_{\mathrm{lin},R_{\mathrm{lin}}, L}^2 \rangle}{\langle \delta_{\mathrm{lin},R_{\mathrm{lin}}, L}^3 \rangle}\left( 1 \pm \sqrt{1 + 2\delta^* \frac{\langle \delta_{\mathrm{lin},R_{\mathrm{lin}}, L}^3 \rangle}{\langle \delta_{\mathrm{lin},R_{\mathrm{lin}}, L}^2 \rangle^2}} \right)\,,
\end{equation}
\begin{equation}
    \lambda \approx \frac{j^*}{\mathcal{F}_{\mathrm{2D}}^{\ '}(\delta^*)} - \left.\frac{\dd \varphi_{\mathrm{lin},R',L}(j^*)}{\dd R'}\right|_{R'=R_{\mathrm{lin}}}\ \frac{R^2}{2 R_{\mathrm{lin}}}\ .
\end{equation}
However, instead of the above equation we need to solve expressions like
\begin{align}
    &\ \tilde{\varphi}_{\mathrm{cy}, R}\left(\lambda\right) \equiv \lim_{L\rightarrow \infty} \frac{\varphi_{\mathrm{cy}, R, L}\left(L\lambda\right)}{L} \nonumber \\
    \approx&\  -\lim_{L\rightarrow \infty} \min_{\delta, j} \frac{1}{L} s_{L\lambda}(\delta, j)\nonumber \\
    =&\  -\lim_{L\rightarrow \infty} \min_{\delta, j} \frac{1}{L} s_{L\lambda}(\delta, Lj)\nonumber \\
    =&  \lim_{L\rightarrow \infty} \min_{\delta, j} \left( \lambda \mathcal{F}_{\mathrm{2D}}(\delta) - \delta j + \frac{1}{L}\varphi_{\mathrm{lin}, R\sqrt{1+\mathcal{F}_{\mathrm{2D}}(\delta)}, L}(Lj)\right)\,,
\end{align}
In the limit $L \gg R'$ we have
\begin{align}
    \langle \delta_{\mathrm{lin},R', L}^2 \rangle \sim \frac{1}{L}\ ,\ \langle \delta_{\mathrm{lin},R', L}^3 \rangle \sim&\ \frac{1}{L^2}\,,
\end{align}
\begin{align}
    \Rightarrow \varphi_{\mathrm{lin},R', L}(Lj) \sim L\ .
\end{align}
Hence, there is a well-defined limit
\begin{align}
    \tilde{\varphi}_{\mathrm{lin},R'}(j) \equiv&\ \lim_{L\rightarrow \infty} \frac{1}{L}\varphi_{\mathrm{lin},R',L}(Lj) \,,
\end{align}
and the cylindrical CGF becomes
\begin{align}
    &\ \tilde{\varphi}_{\mathrm{cy}, R}\left(\lambda\right) \equiv \lim_{L\rightarrow \infty} \frac{\varphi_{\mathrm{cy}, R, L}\left(L\lambda\right)}{L} \nonumber \\
    \approx&\  \min_{\delta, j} \left( \lambda \mathcal{F}_{\mathrm{2D}}(\delta) - \delta j + \tilde{\varphi}_{\mathrm{lin}, R\sqrt{1+\mathcal{F}_{\mathrm{2D}}(\delta)}}(j)\right)\, .
\end{align}
In practice, this minimisation can be carried out by first fixing a value of $\delta^*$, then defining
\begin{align}
    S_{\mathrm{lin},3,\infty} \equiv&\ \lim_{L \rightarrow \infty} \frac{\langle \delta_{\mathrm{lin},R_{\mathrm{lin}}, L}^3 \rangle}{\langle \delta_{\mathrm{lin},R_{\mathrm{lin}}, L}^2 \rangle^2}\,, \\
    E_{\mathrm{lin},3,\infty} \equiv&\ \lim_{L \rightarrow \infty} \frac{1}{L} \frac{\langle \delta_{\mathrm{lin},R_{\mathrm{lin}}, L}^2 \rangle}{\langle \delta_{\mathrm{lin},R_{\mathrm{lin}}, L}^3 \rangle}\,,
\end{align}
and then solving
\begin{equation}
    j^* = E_{\mathrm{lin},3,\infty}\left( 1 \pm \sqrt{1 + 2\delta^* S_{\mathrm{lin},3,\infty}} \right) \,,
\end{equation}
\begin{equation}
    \lambda \approx  \frac{j^*}{\mathcal{F}_{\mathrm{2D}}^{\ '}(\delta^*)} - \left.\frac{\dd \tilde{\varphi}_{\mathrm{lin},R'}(j^*)}{\dd R'}\right|_{R'=R_{\mathrm{lin}}}\ \frac{R^2}{2 R_{\mathrm{lin}}}\ .
\end{equation}
The Limber approximation for the CGF then becomes
\begin{equation}
    \varphi_{w,\theta}(\lambda) \approx \int \dd w\ \tilde{\varphi}_{\mathrm{cy}, \chi\theta}\left(w(\chi)\lambda, \chi\right)\ .
\end{equation}
If we consider two different projections of the density field with projection kernels $w_1(\chi)$ and $w_2(\chi)$, and if both of these projected fields are smoothed by the same angular scale $\theta$, then their joint CGF is similarly approximated by
\begin{equation}
\label{eq:projection_integral}
    \varphi_{w_1, w_2,\theta}(\lambda_1, \lambda_2) \approx \int \dd \chi\ \tilde{\varphi}_{\mathrm{cy}, \chi\theta}\left([w_1(\chi)\lambda_1 + w_2(\chi)\lambda_2], \chi\right)\ .
\end{equation}

\subsection{Practical evaluation of inverse Laplace transformations}
\label{sec:invLaplace}
In order to successfully employ FFT, the evaluation of $\varphi_\theta(i\lambda_\delta, i\lambda_\kappa)$ must be carried out on a grid of imaginary values $i\lambda_\delta$ and $i\lambda_\kappa$, which requires a complex continuation of $\varphi_\theta(i\lambda_\delta, i\lambda_\kappa)$. In practice, we do this through a complex continuation of the integrand in \eqnref{projection_integral}. This is more straightforward, because at each integration step $w$ the function $\tilde{\varphi}_{\mathrm{cy}, \chi\theta}(\cdot\, ,\, w)$ is only a CGF of one variable, as both the galaxy density and weak lensing convergence are predicted from the underlying matter density.

To continue such a 1-dimensional CGF to the complex $\lambda$-plane, one might \eg try to approximate $\tilde{\varphi}_{\mathrm{cy}, \chi\theta}(\lambda\, ,\, w)$ with a finite power series in $\lambda\,$. But this is numerically unstable, since typical large-scale structure CGFs display a branch cut in the complex $\lambda$ plane \citep{Bernardeau2000, Valageas2002, Friedrich2020}. To avoid this instability, we follow the steps described by \cite{Friedrich2020}:
\begin{enumerate}
    \item[A)] Define the auxilliary variable 
    \begin{equation}
        \tau(\lambda) := \mathrm{sign}(\lambda) \cdot \sqrt{2(\lambda \tilde{\varphi}_{\mathrm{cy}, \chi\theta}(\lambda) - \tilde{\varphi}_{\mathrm{cy}, \chi\theta}'(\lambda))}\ .
    \end{equation}
    \item[B)] Fit a polynomial of finite order $N$ in $\tau$ to both $\lambda(\tau)$ and $\varphi_{\mathrm{cy}, \chi\theta}(\lambda(\tau))$. 
    \item[C)] To convert the problem of function composition into matrix multiplication, determine the $N\times N$ Bell-Jabotinsky matrices \citep{Jabotinsky1963} of $\lambda(\tau)$ and  $\varphi_{\mathrm{cy}, \chi\theta}(\lambda(\tau))$ \wrt $\tau$
    \begin{equation}
        \left(B^{\lambda | \tau}\right)_{k,\ell} = \frac{1}{k!} \left. \frac{\dd^k \lambda^\ell}{\dd \tau^k} \right|_{\tau = 0}\ ;\ \left(B^{\varphi | \tau}\right)_{k,\ell} = \frac{1}{k!} \left. \frac{\dd^k \varphi^\ell}{\dd \tau^k} \right|_{\tau = 0}\,.
    \end{equation}
    \item The Bell-Jabotinsky matrix of the convolution of the two polynomials $\phi$ \wrt $\lambda$ (whose column $\ell=1$ contains the cumulants of order $k$ divided by $k!$) is then given by
    \begin{equation}
        \mathbf{B}^{\varphi | \lambda} = \mathbf{B}^{\varphi | \tau} \cdot \left(\mathbf{B}^{\lambda | \tau} \right)^{-1}\ .
    \end{equation}
\end{enumerate}
It was shown by \cite{Friedrich2020} that for typical cases this procedure converges up to the 7th cumulant for a polynomial degree of about $N=16$. \Ie we perform the above steps for $N=16$, but then only use the resulting power series up to the 7th order term for the complex continuation.

\begin{figure*}
    \centering
\includegraphics[width=0.8\textwidth]{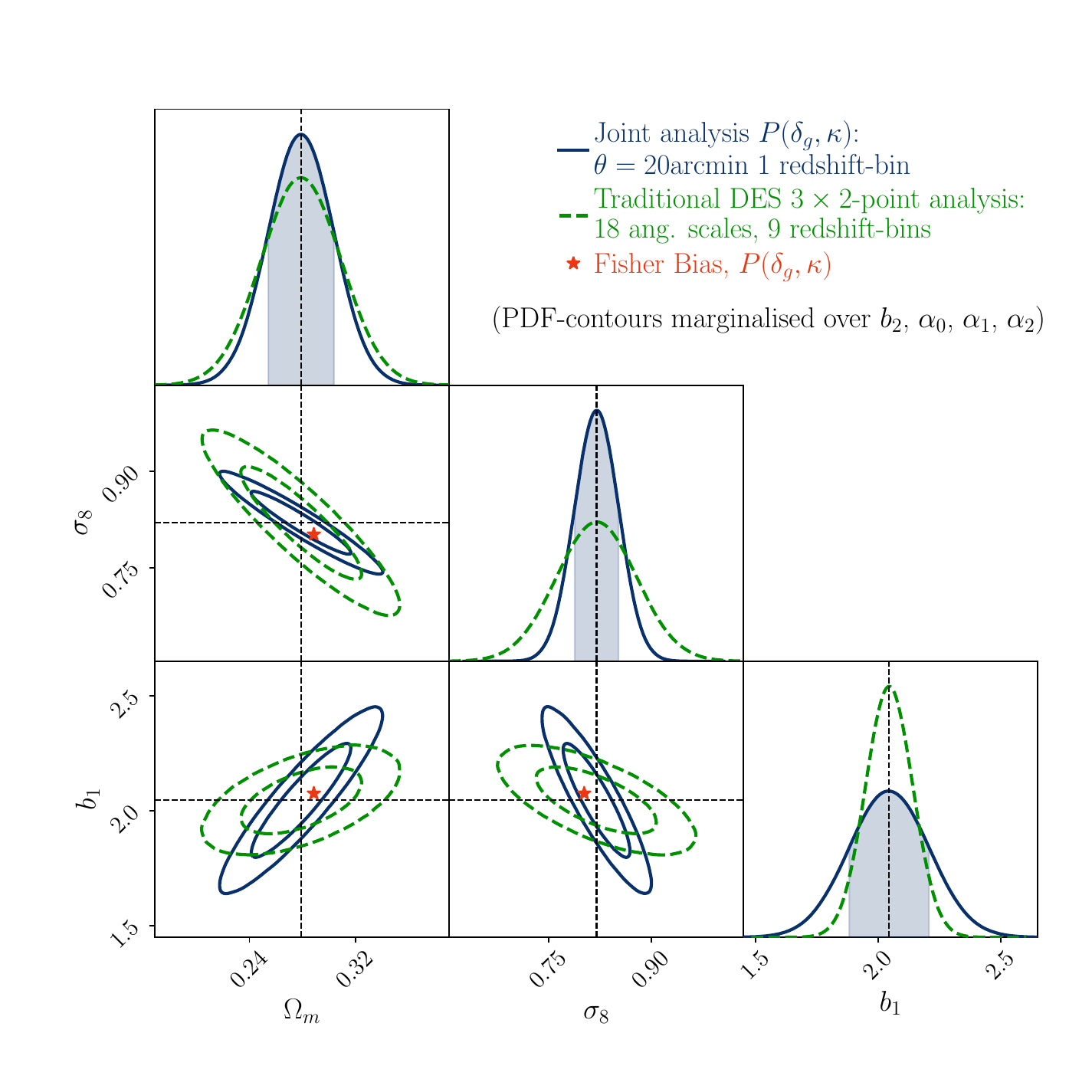}
\caption{Fisher forecast for attainable constraints on cosmological parameters and linear Eulerian bias $b_1$. Solid blue contours assume the PDF-analysis we presented here and are marginalised over quandratic Eulerian bias $b_2$ as well as the three shot-noise parameters $\alpha_0\,$, $\alpha_1$ and $\alpha_2\,$. We compare this with a DES-like, fully tomographic 3$\times$2-point analysis (dashed green contours). The 3$\times$2-point constraints have been marginalised over a number of nuisance parameters, including the linear bias parameters of all lens redshift bins, except the one corresponding to the lens redshift bin that was also used in the PDF analysis (\cf \secref{constraints} for details).}
  \label{fi:forecast_jointPDF_3}
\end{figure*}

\begin{figure*}
    \centering
\includegraphics[width=0.9\textwidth]{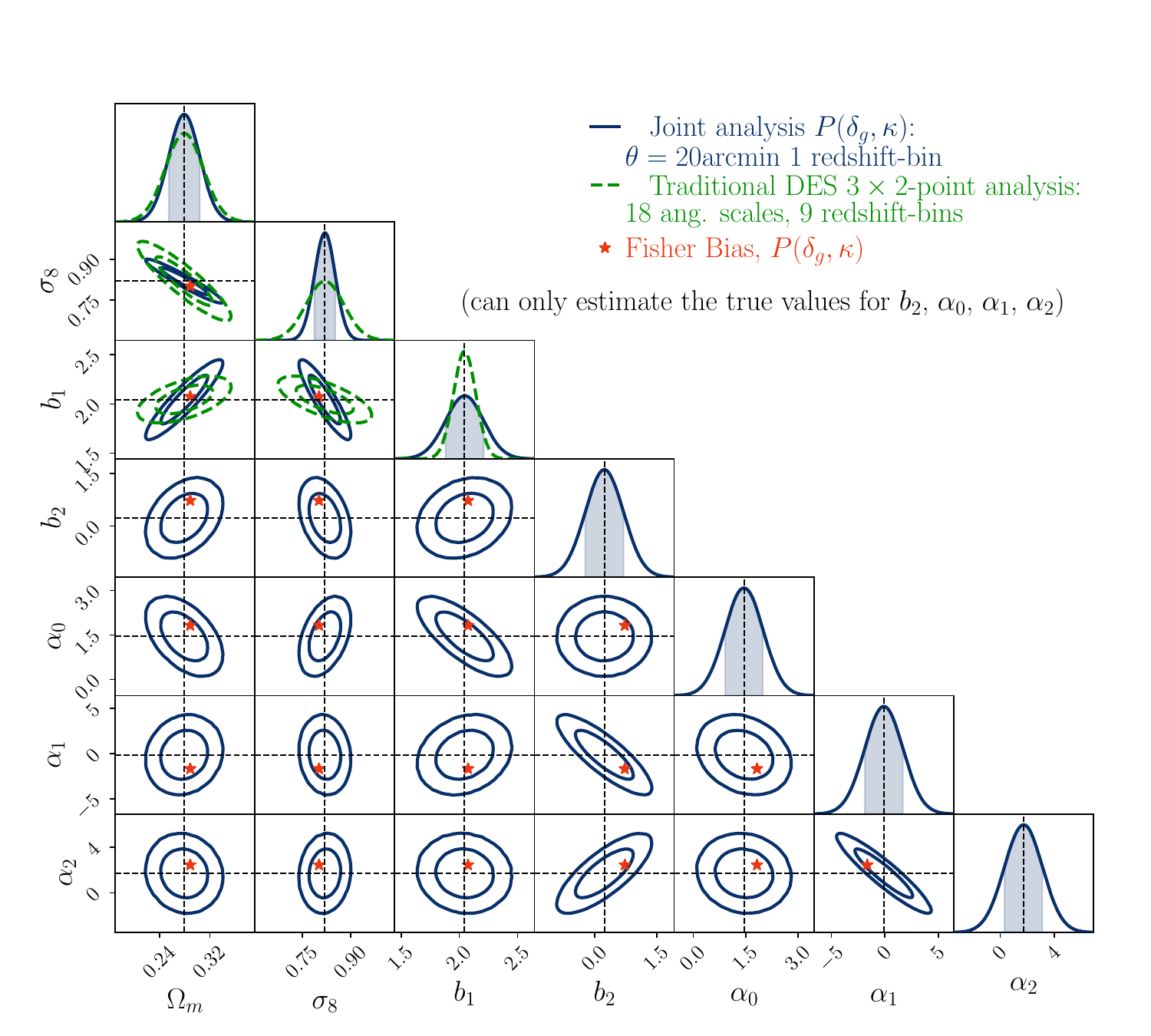}
\caption{Same as \figref{forecast_jointPDF_3}, but now also showing constraints on quandratic bias $b_2$ and the shot-noise parameters $\alpha_0\,$, $\alpha_1$ and $\alpha_2\,$. As explained in \secref{robustness}, we only have a potentially biased estimate of the true values for these parameters. Hence, the parameter bias indicated by the orange markers is likely overestimating the actual bias of a PDF-analysis for these parameters.}
  \label{fi:forecast_jointPDF}
\end{figure*}

\section{Model validation with mock data}
\label{sec:validation}

\subsection{N-body simulations by \citet{Takahashi2017} and HOD-approach for mock galaxy sample}
\label{sec:mock_data}

We use the publicly available cosmological simulation data from Ref.~\cite{Takahashi2017} (hereafter referred to as the T17 simulations) to test our theoretical model for the joint PDF.\footnote{The simulation data products are at \url{http://cosmo.phys.hirosaki-u.ac.jp/takahasi/allsky_raytracing/}.} In our work, we use the full-sky lightcone halo catalogues and cosmic shear lensing maps of the simulation suite. The simulation data products were obtained from a gravity-only N-body simulation in a $\Lambda$CDM cosmology with the following parameters : $\Omega_{\rm cdm} = 0.233,\; \Omega_{\rm b} = 0.046,\; \Omega_{\rm m} = \Omega_{\rm cdm} + \Omega_{\rm b} = 0.279,\; \Omega_{\rm \Lambda} = 0.721,\; h = 0.7,\; \sigma_8 = 0.82\; \mathrm{and}\; n_s = 0.97$. For all calculations in this paper. we adopt this as our fiducial cosmology. Halos and sub-halos in the simulation were identified using the six-dimensional phase-space friends-of-friends algorithm \verb|ROCKSTAR| \citep{Behroozi2013}. These halo catalogues were combined in layers of shells to obtain full-sky lightcone halo catalogues. The simulation boxes were also ray traced using the multiple-lens plane ray-tracing algorithm \verb|GRAYTRIX| \citep{Hamana2015, Shirasaki2015} to obtain weak lensing convergence maps. We utilize 108 realizations of these data products, obtained from multiple realizations of the T17 simulations all of which come in \verb|Healpix| \citep{Zonca2019} format using the $n_{\rm side}=4096$ resolution.

As discussed in Refs.~\cite{Takahashi2017, Shirasaki_2019}, the T17 maps suffer from systematic effects associated with, for example, the thickness of the lens shells that may need to be incorporated into the theory predictions for fairer comparisons. In this work, we do not perform these corrections explicitly since, as we will see below, we find good agreement between simulations and our theory predictions.

\subsubsection{Simulated weak lensing convergence map}

The T17 simulation suite provides weak lensing maps for $N=38$ fixed source redshift planes between $z_s = 0.05$ and $z_s = 5.3$. We combine these maps according to a source distribution $n_s(z)$ inspired by a predicted fourth tomographic bin of the Euclid Year 1 analysis as explained in \apprefnospace{source_n_z} (see Fig. \ref{fig:redshift_distributions}). The procedure to do that is as follows:
\begin{equation}
    \kappa_{n(z)} = \sum_{i=1}^{N} s_i \kappa_{z_{s}^i} \ ,
\end{equation}
where $s_i$ is a specific weight for a given full-sky lensing map $\kappa_{z_{s}^i}$ at source plane $z_{s}^i$. This final simulated convergence map can then be expressed as a line-of-sight projection of the matter density contrast through equation \eqref{eq:def-convergence} with a lensing kernel $w_{n_s(z)}$:
\begin{equation} 
    w_{n(z)}(z) = \sum_{i=1}^{N} s_i w_{z_{s}^i}(z) \,,
\end{equation}
where $w_{z_{s}^i}$ is given by equation~\eqref{eq:lensing_kernel_zs}. This is a discretisation of equation~\eqref{eq:lensing_kernel}, with the weights given by
\begin{equation}
    s_i
    =\Delta z^i_s \int_{0}^1 d\mu\, n_s\left(z_s^{i-1}+ \mu \Delta z_s^{i}\right) \,.
\end{equation}
The corresponding discretised redshift distribution is shown as the blue histogram in Figure~\ref{fig:redshift_distributions}.

For our analysis, we generate convergence maps with and without shape noise. To mimic Euclid-like shape noise, we add to every pixel of the final noiseless convergence map Gaussian noise with zero-mean and variance 
\begin{equation}
    \sigma_{N}^2 = \frac{\sigma_e^2}{n_g \ \cdot A_{pix}} \ ,
\end{equation}
where $A_{pix}$ is the area of the pixel at the given \verb|NSIDE| of the \verb|Healpix| map, $\sigma_e = 0.26$ is the dispersion of intrinsic galaxy ellipticities
as found for weak lensing surveys, and $n_g$ is the number density of galaxies that we assume to be $n_g=6$arcmin$^{-2}$ as expected for the fourth tomographic source redshift bin of Euclid.

\subsubsection{Mock HOD galaxy catalogues}

Being a gravity-only N-body simulation, the T17 suite does not come with galaxy catalogues. We hence use the all-sky mock HOD map from  Ref.~\cite{Halder_2023} who populated the T17 halo catalogues using an empirical Halo Occupation Distribution (HOD) method \citep{Berlind_2002, Cooray2002} to mimic the MagLim-like third redshift bin of DESY3 \citep{DES:2021bpo}. We refer the reader to \cite{Halder_2023} for details about the creation of the HOD map. The shape of the mock galaxy distribution in the redshift bin is shown with the red line in Fig.~\ref{fig:redshift_distributions}.

\subsubsection{Measurements: data vector and covariance matrix}

We measure the joint one-point galaxy and convergence PDF from 6 non-overlapping 5000 ${\rm deg}^2$ circular footprints carved from each all-sky T17 galaxy/convergence map. Over the 108 T17 realizations, this results in a total of 648 galaxy and convergence footprints that we use to obtain our mean data vector and to estimate its covariance. For our subsequent Fisher analysis, we use the galaxy density contrast $\delta_g$ instead of the galaxy count $N_g$. We obtain the galaxy density contrast maps by computing the mean number density and using \eqnref{delta_g} in each of the footprints.

We compute smoothed convergence and galaxy density contrast maps using a circular top-hat filter with radius of $20$arcmin by applying a window function in spherical harmonic space with a circular beam profile. For this, we use the \texttt{healpy.sphtfunc.beam2bl} function from \texttt{Healpix}\footnote{\url{https://healpy.readthedocs.io/en/stable/index.html}}, with a maximum multipole moment  $\ell_{\rm max}=3\cdot$\verb|NSIDE|$-1$.  This method is computationally more efficient than smoothing in angular space by querying neighbouring pixels and averaging over them. This is in particular relevant for the extraction of overlapping apertures in a high map resolution that reduces finite sampling errors and thus the covariance \citep{Uhlemann2023PDFcov}.

The joint PDF from a single footprint is extracted as 2D histograms of the smoothed galaxy density contrast and the convergence maps. The full mean data vector is then obtained as the average over the estimates from the 648 mock footprints.

For our Fisher analysis, we rebin the theoretical and simulated joint PDFs onto a coarser 2D grid with 10 bins along each axis. We then retain only the region containing $95\%$ of the total probability, which corresponds to the $2\sigma$ bulk of the probability distribution where the theoretical predictions are reliable. After cutting the outermost $5\%$ of the probability (i.e., the tails), we flatten the remaining 2D joint PDF into a 1D array of length $N_d = 58$.

The data covariance matrix is estimated from the mocks as
\begin{equation}
    \label{eq:cov_estimation}
    \mathbf{\hat{C}}= \frac{1}{N_s - 1}\sum_{i=1}^{N_s}(\hat{d}_i - \hat{d})(\hat{d}_i - \hat{d})^{\rm T},
\end{equation}
where $N_s = 648$ is the number of mock footprints, $\hat{d}_i$ is the data vector measured in the $i$-th footprint, and $\hat{d}$ the sample mean over the $N_s$ realizations. To get an unbiased estimate of the inverse covariance matrix, we apply the correction \citep{Hartlap2007}
\begin{equation}
    \label{eq:precision_hartlap}
    \mathbf{C}^{-1} = \frac{N_{\rm s} - N_{\rm d} - 2}{N_{\rm s} - 1}\mathbf{\hat{C}}^{-1} \  .
\end{equation}

\subsection{Comparison of PDF model and simulations}
In Figure~\ref{fi:fiducial_PDF} (a) we show a comparison of the theoretical prediction (contour lines) and the simulation result (colour map) for the joint PDF of the weak lensing convergence $\kappa$ and photometric galaxy number count $N_g$ in apertures of 20arcmin. The theory model is computed using galaxy bias and stochasticity parameters $\{b_1^E,b_2^E,\alpha_0,\alpha_1,\alpha_2\}$ obtained from fitting the conditional mean and variance of galaxy counts given dark matter density following equations~\eqref{eq:bias_model_Eulerian}~and~\eqref{eq:stochasticity_model} along with the measured lens number density $n_g^{\rm 3D}=1.2 \times 10^{-3}$ Mpc$^{-3}$ corresponding to an angular density of $n_g=0.0834$ arcmin$^{-2}$. Figure~\ref{fi:fiducial_PDF} (b) illustrates the corresponding residuals in the joint PDF and the corresponding marginals. We find an accuracy at the $\{1,2,4,16\}\%$-level in the central region of the joint PDF containing $\{12,34,68,95\}\%$ of probability, respectively. 

\subsection{Response to changing physical parameters}

In Figure~\ref{fi:derivatives}, we demonstrate how the joint PDF of the weak lensing convergence and galaxy density contrast responds to changing the underlying parameters. We consider cosmology ($\sigma_8$ and $\Omega_m$) in the top row, Eulerian galaxy bias ($b_1^E$ and $b_2^E$) in the middle row, and non-Poisson galaxy stochasticity ($\alpha_0$ and $\alpha_1$) in the bottom row.

Increasing the fluctuation amplitude $\sigma_8$ leads to a simultaneous increase in the variances of $\kappa$ and $\delta_g$ through its impact on the matter variance, such that the peak is lowered and the flanks are increased. In contrast, changing the matter density $\Omega_m$ increases the $\kappa-$variance, but decreases the $\delta_g$-variance. In the context of $\Lambda$CDM and the linear regime, the matter variance is proportional to the square of the growth factor, $D(z) \propto \Omega_m^{-1/2}$. Consequently, increasing $\Omega_m$ reduces the $\delta_g-$variance, while the $\kappa-$variance is dominated by the weak lensing kernel, which is proportional to $\Omega_m$.  It flattens the conditional mean $\langle \delta_g|\kappa\rangle$, similar to decreasing the linear bias. This is because increasing $\Omega_m$ while keeping $\sigma_8$ fixed requires smaller linear density fluctuations.

As expected, changing the galaxy bias parameters only affects the $\delta_g$-direction. Increasing the linear Eulerian bias $b_1$ leads to an increased $\delta_g$-variance and thus a decrease of the PDF peak along with an increase in the flanks. In contrast, the quadratic Eulerian bias $b_2$ leads to an antisymmetric signal around the peak.

Varying the galaxy stochasticity parameters changes the joint PDF  mainly in the direction perpendicular to the conditional mean $\langle\delta_g|\kappa\rangle$, which is largely determined by the scatter around the conditional mean $\langle(\delta_g-\langle\delta_g|\delta_m\rangle)^2|\delta_m\rangle$. Increasing the non-Poissonian amplitude $\alpha_0$ spreads the joint PDF in a direction perpendicular to the conditional mean. In contrast, increasing the linear density-dependence of the stochasticity, $\alpha_1$, leads to an antisymmetric signal around the diagonal direction. Changing the quadratic term, $\alpha_2$, leads to a symmetric signature with an additional zero-crossing (not shown).

\subsection{Cosmological power of the joint PDF compared to 3$\times$2-point}
\label{sec:constraints}

To get a qualitative sense of the constraining power of the joint PDF $p(\delta_g, \kappa)$ we perform a Fisher forecast. This means that we derive the width of the parameter constraints that an analysis of $p(\delta_g, \kappa)$ would yield if its statistical uncertainties were Gaussian with a parameter-independent covariance matrix and if the shape of $p(\delta_g, \kappa)$ depended only linearly on the parameters $(\Omega_m, \sigma_8, b_1, b_2, \alpha_0, \alpha_1, \alpha_2)$.

The solid, blue contours in \figref{forecast_jointPDF} show the 1-$\sigma$ and 2-$\sigma$ constraints predicted for an analysis of the pdf bulk (\ie when cutting away the outermost $5\%$ at the flanks of $p(\delta_g, \kappa)$). The covariance matrix assumed for this forecast is the one estimated from the simulated data described in \secref{mock_data}. Note that even though we use 5 parameters to describe the relation between galaxies and matter (\cf \secref{cylinder_CGF_with_bias} and \secref{shot-noise}), such a PDF analysis can still yield meaningful constraints on the cosmological parameters $\Omega_m$ and $\sigma_8$. This is exactly the richness offered by analysing the full shape of the joint PDF $p(\delta_g, \kappa)$. As we had demonstrated in \figref{derivatives}, changing different model parameters causes complex and varied signatures in the $\delta_g$-$\kappa$ plane that can be easily distinguished from each other.

For comparison, we also display corresponding constraints for a 2-point function analysis with the green dashed contours. In practice, we assume a Fisher matrix corresponding to the Year-3 analysis of the 3$\times$2-point functions in the Dark Energy Survey \citep{DES:2021wwk} - in particular, we assume the fiducial analysis setup that was considered in \cite{Friedrich2021cov}. Note that while this 3$\times$2-point setup is more realistic than our PDF setup in certain aspects, it is less realistic in others. On the one hand, the Fisher contours of \cite{Friedrich2021cov} include marginalisation over a large set of nuisance parameters such as intrinsic source alignment, photometric redshift uncertainties and multiplicative shear biases (though the latter two are constrained by tight priors). Our PDF-based forecast does not include these nuisance effects. On the other hand, the DES year-3 fiducial analysis marginalises over only one parameter for the galaxy-matter connection, namely a single linear bias parameter for each redshift bin, while our PDF-based analysis constrains a much more nuanced view of this relation with overall 5 different parameters. Despite this complex model for the galaxy matter connection, our assumed PDF analysis can still measure $\Omega_m$ and $\sigma_8$ with similar uncertainties as the 3$\times$2-point analysis. This is particularly impressive, because we compare to a tomographic 3$\times$2-point analysis with 5 lens redshift bins, 4 source redshift bins, and measurements across 18 different angular scales. In contrast, the PDF analysis only considers cosmological fields in a single lens and source redshift bin combination and is measured only at one angular scale.

To understand this surprising (albeit qualitative) comparison, we find it instructive to think of the pair (PDF, 2-point functions) as follows:
\begin{itemize}
    \item 2-point correlation functions measure the 2nd moment of the cosmic density field as a function of scale;
    \item The PDF measures (in principle) all moments of the density field at one scale.
\end{itemize}
While the 3$\times$2-point functions capture the auto- and cross-correlations, the joint PDF captures mixed higher-order moments. From results for the matter and weak lensing PDF \citep{Uhlemann2023PDFcov}, we expect that the cross-covariance between the joint PDF and the 3$\times$2-point data vector will largely stem from the correlation between the 2-point functions and the variance of the PDF.
Of course, if we were to perform a combined analysis of the PDF measured at various smoothing scales and for sources and lenses at different redshift ranges, then the scale- and redshift-dependent information of the 2-point function would be completely contained within the PDF analysis. However, we expect that it will be more practical (and at little cost to the overall constraining power) to simply perform a combined analysis of the joint PDF at one scale and the 3$\times$2-point data vector. Due to the complementarity of the contours in \figref{forecast_jointPDF}, it seems that such a combined analysis would indeed be very beneficial, but we leave a full exploration of such combined analyses for future work.

\subsection{Robustness of parameter constrains}
\label{sec:robustness}

While \figref{fiducial_PDF} indicates that the relative deviation between our PDF model and measurements in simulated data is small, it is not immediately clear which impact these (albeit small) residuals would have on parameter constraints. To obtain an estimate of this effect we calculate the \emph{Fisher bias} between the simulated data and our model. This means that we consider a linear Taylor expansion of our model around the fiducial parameters of the simulations and then find the parameters for which this linearized models fits the simulated data best (see \appref{Fisher_bias} for details). 

The orange stars in \figref{forecast_jointPDF_3} show the location of these linearized best-fitting parameters, and the difference between those and the true parameters of the simulations gives an estimated of how biased parameter constraints of DES-like data with our model would be. You can see in the figure, that for $\Omega_m\,$, $\sigma_8$ and $b_1$ this bias is well within the expected 1-$\sigma$ uncertainties of such an analysis. In \figref{forecast_jointPDF} we also show the Fisher biases for the remaining parameters $b_2$ as well as $\alpha_0\,$, $\alpha_1$ and $\alpha_2\,$. One problem that we face here is that - in contrast to $\Omega_m$ and $\sigma_8$ - we do not a-priori know the true values for our bias and shot-noise parameters in our simulated data. To estimate these parameters, we compare maps our simulated galaxy density to maps of simulated matter density. Projected matter density in the simulations of \cite{Takahashi2017} is available in concentric shells, and a weighted sum of such shells can only approximate the lens redshift distribution that is displayed in Figure~\ref{fig:redshift_distributions}. To account for this, we create a second galaxy sample that has the same step-like redshift distribution as the stacked matter density shells. Comparing density fluctuations between the resulting matter density and galaxy density maps allows us to directly measure the conditional expectation value $\langle \delta_g | \delta_m\rangle$ and the conditional variance $\mathrm{Var}(N_g | \delta_m)$. We then fit our bias and shot-noise parameters to these measured functions to obtain estimates of their true value. We expect that this is robust for linear bias $b_1$, as long at the redshift distriution of the two density maps roughly matches our target distribution. But - as was \eg shown in \cite{Friedrich2018, Friedrich2021b}, the other parameters of our bias and shot-noise model are very sensitive to the details of the considered galaxy sample. Thus, the Fisher biases displayed in \figref{forecast_jointPDF} are likely overestimating the actual bias of a PDF-analysis for these parameters. This is also supported by the fact that the cosmological parameters are recovered very well, even after marginalising over our complete bias and shot-noise model.

\section{Discussion \& conclusions}
\label{sec:discussion}

In this work, we developed a theoretical framework to model the joint probability distribution of galaxy clustering number counts and the weak lensing convergence. This was done by incorporating the connection between dark matter and galaxy counts through an Eulerian galaxy bias, and accounting for stochasticity using a quadratic shot noise model as a function of the underlying matter density contrast. We also convolve the impact of Gaussian shape noise into the weak lensing convergence which enters the joint PDF.

We validate our theoretical prediction for the joint PDF $p(N_g, \kappa)$ by comparing it with measurements on convergence maps and HOD catalogues built from the T17 simulation suite \cite{Takahashi2017}. The weak lensing convergence maps have a redshift distribution resembling the fourth of 5 equipopulated redshift bins of a Stage-IV-like survey such as Euclid. The mock HOD catalogs mimic the third redshift bin of the MagLim sample of DES Y3 \cite{DES:2021bpo}. We find good agreement of the order of $5\%$ in the bulk of the joint distribution, within the central $69\%$ of probability, and differences up to about $16\%$ at the edges of the central $95\%$ probability contour.

A simple Fisher analysis for the cosmological parameters $\Omega_m$ and $\sigma_8$, the galaxy bias parameters ${b_1, b_2}$, and the shot noise parameters ${\alpha_0, \alpha_1, \alpha_2}$ demonstrates the strong constraining power of the joint PDF, allowing these parameters to be simultaneously constrained with meaningful precision. Compared to a full 3$\times$2-point analysis, the joint PDF using only a single lens and source redshift bin and a single angular smoothing scale yields comparable uncertainties. This makes the joint PDF a promising tool for cosmological analyses beyond two-point statistics, enabling efficient extraction of non-Gaussian information.
Future work will focus on incorporating additional sources of systematics, such as proper forward modelling of the lensing convergence reconstruction from the shear components \cite{Barthelemy2024, Castiblanco2024}, intrinsic alignments, baryonic feedback, and photometric redshift uncertainties.  A complete analysis will require a tomographic approach that combines multiple lens and source redshift bins, which has demonstrated strong potential for constraining cosmological parameters in a weak lensing PDF analysis \citep{Castiblanco2024}, as well as in a combined analysis with 3$\times$2-point statistics.

While we focused on the weak gravitational lensing of galaxy shapes here, the formalism can be applied to CMB lensing when incorporating post-Born effects \citep{Barthelemy2020postBorn} in the theoretical model. 

\section*{Acknowledgements}

We would like to thank Francis Bernardeau, Daniel Gruen, Alexandre Barthelemy, Sandrine Codis and Alex Gough for fruitful discussions. OF is grateful to LC, AH and CU for breathing life into this project, when it was dormant. Particular thanks goes to LC, who shaped large parts of this manuscript.  OF was supported by a Fraunhofer-Schwarzschild-Fellowship at Universit\"atssternwarte M\"unchen (LMU observatory) and by DFG's Excellence Cluster ORIGINS (EXC-2094 – 390783311). LC and CU were supported by the STFC Astronomy Theory Consolidated Grant ST/W001020/1 from UK Research \& Innovation. CU was also supported by the European Union (ERC StG, LSS\_BeyondAverage, 101075919). This work was supported by the Deutsche Forschungsgemeinschaft (DFG, German Research Foundation) via the PaNaMO project (project number 528803978) and by the Munich Institute for Astro-, Particle and BioPhysics (MIAPbP), which is funded by the Deutsche Forschungsgemeinschaft (DFG, German Research Foundation) under Germany's Excellence Strategy – EXC-2094 – 390783311.
The figures in this work were generated using  Matplotlib
\cite{Hunter:2007} making use of the NumPy \cite{harris2020array} and ChainConsumer \cite{Hinton2016} Python libraries. We are indebted to Stella Seitz for her wide-ranging contributions to the field of gravitational lensing, one of which is at the centre of the present work. May she rest in peace.

\appendix

\section{Cylindrical collapse in $\Lambda$CDM}
\label{app:spherical_collapse}

In the Newtonian approximation and geometrised units $G = 1 = c$ the evolution of cylindrical perturbations $\delta$ is described by
\begin{equation}
\label{eq:SC_in_conformal_time}
\ddot{\delta} + \mathcal{H} \dot{\delta} - \frac{3}{2} \frac{\dot{\delta}^2}{1+\delta} \ = 4\pi \bar \rho_m a^2 \delta (1+\delta)\ ,
\end{equation}
where $\tau$ is conformal time, $\mathcal{H} = \dd \ln a / \dd \tau$ is the conformal expansion rate (see \cite{Friedrich2020} for a general expression following \citet{MukhanovBook} for the planar and spherical cases). To evaluate the cylindrical collapse \eqnref{delta_R_equals_F} and related expressions we  solve \eqnref{SC_in_conformal_time} with the initial conditions
\begin{equation}
    \delta_i = \delta_{L,R_L}^*\ D(z_i)\ ,\ \dot{\delta_i} = \delta_i\ \mathcal{H}(z_i)\ ,
\end{equation}
where $z_i$ is a redshift chosen during matter domination. (In fact, in our calculation of $D(z)$ we set the radiation density $\Omega_r$ to zero and then choose $z_i = 4000$.)

\section{Source galaxy redshift distribution}
\label{app:source_n_z}

We assume a model of the true source galaxy redshift distribution for Stage-IV survey-like given by \cite{Fu:2007qq}
\begin{equation}
n_s^{true}(z) = A \frac{z^a+z^{ab}}{z^b+c}\,.
\end{equation}
with $A = 1.7865$arcmin$^{-2}$, $a=0.4710$, $b=5.1843$, $c=0.7259$, \cite{Martinet:2020mqm}. 

To account for photometric redshift uncertainties, we assume that the probability distribution for the observed photometric redshift given the true galaxy redshift follows a Gaussian distribution \cite{LSSTDarkEnergyScience:2018jkl}, as follows
\begin{equation}
n(z_{ph}|z) = \frac{1}{\sqrt{2\pi\sigma^2(z)}}\exp\left[-\frac{1}{2}\frac{(z-z_{ph})^2}{\sigma^{2}(z)}\right]\,,     
\end{equation}
with $\sigma(z) = 0.05(1+z)$.
The source distribution in a tomographic bin is then given by
\begin{equation}
    n^{i}(z_{ph}) =  \int^{z_{max}}_{z_{min}}\, dz \,n(z_{ph}|z)*n_s^{i,true}(z)\,.
\end{equation}

\section{Model bias from Fisher matrix}
\label{app:Fisher_bias}

Parameter biases $\delta\theta$ arising from offsets between measured and predicted statistics, $\Delta S = S^{\rm meas} - S^{\rm pred, fid}$, can be computed using the Fisher formalism. Maximizing the likelihood yields
\begin{equation}
\label{eq:Fisher_bias}
    \delta\theta_i=(F^{-1}_{\rm pred})_{ij} \frac{\partial S_\alpha^{\rm pred}}{\partial \theta_j} C^{-1}_{\alpha\beta} \Delta S_\beta\,.
\end{equation}
This expression is derived by minimizing the $\chi^2$ function
\begin{subequations}
\begin{equation}
\label{eq:chi2_alt}
    \chi^2(\theta) = \frac{1}{2} \left(S_\alpha^{\rm pred}(\theta)-S_{\alpha}^{\rm meas}\right) C_{\alpha\beta}^{-1} \left(S_{\beta}^{\rm pred}(\theta)- S_{\beta}^{\rm meas}\right)\,,
\end{equation}  such that
\begin{equation}
\frac{\partial\chi^2}{\partial\theta_i} (\hat\theta) =  \frac{\partial S_\alpha^{\rm pred}}{\partial \theta_i }\Big|_{\hat\theta} C^{-1}_{\alpha\beta} \left(S_\beta^{\rm pred}(\hat\theta)-S_\beta^{\rm meas}\right) \stackrel{!}{=} 0 \,,
\end{equation}
assuming a parameter-independent covariance.
Expanding $S^{\rm pred}$ around the fiducial parameters $\theta^{\rm fid}$ and defining $\hat\theta_i = \theta^{\rm fid}i + \delta\theta_i$, we write
\begin{equation}
    S^{\rm pred}_\beta(\hat\theta) \approx S^{\rm pred,fid}_\beta + \frac{\partial S^{\rm pred}_\beta}{\partial\theta_j}\Big|_{\theta^{\rm fid}}\delta \theta_j 
\end{equation}
And since we assume the model to be linear in the parameters, we have that the derivatives at the fiducial parameter and the most likely parameters are the same
\begin{equation}
    \frac{\partial S^{\rm pred}_\beta}{\partial\theta_j}\Big|_{\hat\theta}=\frac{\partial S^{\rm pred}_\beta}{\partial\theta_j}\Big|_{\theta^{\rm fid}}
\end{equation}

Plugging this in we find 
\begin{equation}
\frac{\partial S_\alpha^{\rm pred}}{\partial \theta_i }\Big|_{\theta^{\rm fid}} C^{-1}_{\alpha\beta} \left(\frac{\partial S^{\rm pred}_\beta}{\partial\theta_j}\Big|_{\theta^{\rm fid}}\delta \theta_j -\Delta S\right) \stackrel{!}{=} 0 \,,
\end{equation}
\end{subequations}
and evaluating it at first order in $\delta \theta_i$, where the derivatives are at the fiducial parameters,
thus obtaining
\begin{equation}
 F_{ij}\delta\theta_j = \frac{\partial S_\alpha^{\rm pred}}{\partial \theta_i }\Big|_{\theta^{\rm fid}} C^{-1}_{\alpha\beta} \Delta S \,.
\end{equation}
Solving this leads to
equation~\eqref{eq:Fisher_bias}.

\def\aj{AJ}%
\def\araa{ARA\&A}%
\def\apj{ApJ}%
\def\apjl{ApJ}%
\def\apjs{ApJS}%
\def\ao{Appl.~Opt.}%
\def\apss{Ap\&SS}%
\def\aap{A\&A}%
\def\aapr{A\&A~Rev.}%
\def\aaps{A\&AS}%
\def\azh{AZh}%
\def\baas{BAAS}%
\def\jrasc{JRASC}%
\def\memras{MmRAS}%
\def\mnras{MNRAS}%
\def\pra{Phys.~Rev.~A}%
\def\prb{Phys.~Rev.~B}%
\def\prc{Phys.~Rev.~C}%
\def\prd{Phys.~Rev.~D}%
\def\pre{Phys.~Rev.~E}%
\def\prl{Phys.~Rev.~Lett.}%
\def\pasp{PASP}%
\def\pasj{PASJ}%
\def\qjras{QJRAS}%
\def\skytel{S\&T}%
\def\solphys{Sol.~Phys.}%
\def\sovast{Soviet~Ast.}%
\def\ssr{Space~Sci.~Rev.}%
\def\zap{ZAp}%
\def\nat{Nature}%
\def\iaucirc{IAU~Circ.}%
\def\aplett{Astrophys.~Lett.}%
\def\apspr{Astrophys.~Space~Phys.~Res.}%
\def\bain{Bull.~Astron.~Inst.~Netherlands}%
\def\fcp{Fund.~Cosmic~Phys.}%
\def\gca{Geochim.~Cosmochim.~Acta}%
\def\grl{Geophys.~Res.~Lett.}%
\def\jcap{JCAP}%
\def\jcp{J.~Chem.~Phys.}%
\def\jgr{J.~Geophys.~Res.}%
\def\jqsrt{J.~Quant.~Spec.~Radiat.~Transf.}%
\def\memsai{Mem.~Soc.~Astron.~Italiana}%
\def\nphysa{Nucl.~Phys.~A}%
\def\physrep{Phys.~Rep.}%
\def\physscr{Phys.~Scr}%
\def\planss{Planet.~Space~Sci.}%
\def\procspie{Proc.~SPIE}%

\bibliographystyle{mnras}
\bibliography{literature}

\end{document}